\documentclass[journal,onecolumn]{IEEEtran}
\IEEEoverridecommandlockouts
\newcommand{\tabitem}{~~\llap{\textbullet}~~}
\usepackage{cite}
\usepackage{amsmath,amssymb,amsfonts}
\usepackage{algorithmic}
\usepackage[caption=false,font=footnotesize]{subfig}
\usepackage{graphicx}
\usepackage{textcomp}
\usepackage{xcolor}
\usepackage{color,soul}
\usepackage{hyperref}
\usepackage{comment}
\usepackage[center]{caption}
\sethlcolor{yellow}

\def\BibTeX{{\rm B\kern-.05em{\sc i\kern-.025em b}\kern-.08em
    T\kern-.1667em\lower.7ex\hbox{E}\kern-.125emX}}
\begin{document}

\title{Transitioning towards fit-for-purpose Public Health Surveillance Systems}

\author{\IEEEauthorblockN{Maria~N.~Anastasiadou\IEEEauthorrefmark{1}, Philippos~Isaia\IEEEauthorrefmark{1}, Panayiotis~Kolios\IEEEauthorrefmark{1}, Christos Charalambous\IEEEauthorrefmark{4}}

\IEEEauthorblockA{\IEEEauthorrefmark{1}KIOS Center of Excellence for Research and Innovation, University of Cyprus,
\newline\IEEEauthorrefmark{4}Unit for Surveillance and Control of Communicable Diseases, Ministry of Health}
% \IEEEauthorblockA{\IEEEauthorrefmark{2}Unit for Surveillance 
% and Control of Communicable Diseases, Ministry of Health, 
%Nicosia, Cyprus%\\
%Email: anastasiadou.maria@ucy.ac.cy, isaia.philippos@ucy.ac.cy,  kolios.panayiotis@ucy.ac.cy
}

\maketitle

\begin{abstract}
The COVID-19 pandemic has exposed several weaknesses in the public health infrastructure, including supply chain mechanisms and public health ICT systems. The expansion of testing and contact tracing has been key to identifying and isolating infected individuals, as well as tracking and containing the spread of the virus. Digital technologies, such as telemedicine and virtual consultations, have experienced a surge in demand to provide medical support while minimizing the risk of transmission and infection. The pandemic has made it clear that cooperation, information sharing, and communication among stakeholders are crucial in making the right decisions and preventing future outbreaks. Redesigning public health systems for effective management of outbreaks should include five key elements: disease surveillance and early warning systems, contact tracing and case management, data analytics and visualization, communication and education, and telemedicine. As the world navigates the COVID-19 pandemic, healthcare ICT systems will play an increasingly important role in the future of healthcare delivery. In a post COVID-19 world, several ICT strategies should be implemented to improve the quality, efficiency, and accessibility of healthcare services, including the expansion of telemedicine, data analytics and population health management, interoperability, and cybersecurity.

Overall, this report summarises the importance of early detection and rapid response, international cooperation and coordination, clear and consistent communication, investing in public health systems and emergency preparedness, digital technology and telemedicine, and equity and social determinants of health. These lessons demonstrate the need for better preparedness and planning for future crises and the importance of addressing underlying issues to create a more resilient and accessible digital infrastructure.
\end{abstract}

\section{Introduction}

The COVID-19 pandemic has had a significant impact on healthcare systems worldwide. The novel virus, which caused a highly contagious respiratory illness, has led to a surge in demand for medical care, including testing, treatment, and hospitalisation. In response, many countries have implemented measures to increase capacity and minimise the spread of the virus. One of the key measures has been the expansion of testing and contact tracing, which helps to identify and isolate infected individuals, as well as to track and contain the spread of the virus. Quarantine and isolation protocols have also been put in place to separate those who are infected or have been exposed to the virus. Of course personal protective equipment (PPE) especially for healthcare workers has also played an important role in the response. However, digital technologies have been the ones experiencing a sugre in demand, with digital public health tools and virtual consultations being the most prevalent healthcare services to continue providing medical support while also minimising the risk of transmission and infection. 

Hence, a key outcome of the COVID-19 pandemic was that it has highlighted numerous weaknesses in the existing public health infrastructure, including its supply chain mechanisms and public health ICT systems. This has made it difficult for health officials and health workers to effectively make informed decisions and implement measures to control the spread of the virus. Many healthcare facilities have underestimated and found under prepared to handle the surge and lacked adequate tools to properly address the evolving challenges. 

A second key outcome is that it has become evident that cooperation and information sharing among stakeholders, including governments and healthcare institutions, as well as clear and timely communication with the public, is crucial in making the right decisions and communicating the right information for smooth society functioning \cite{torrentina2020}. However, it is still uncertain whether public health ICT systems in any one country have adapted to cope with future outbreaks \cite{khetrapal2020}. 

Evidently, there is an urgent need to redesign public health systems for effective digital management of these outbreaks that will include the following five key elements \cite{Filip2022}:
\begin{enumerate}
    \item \textbf{\textit{Disease Surveillance and Early Warning Systems:}} Real-time monitoring and surveillance of diseases can help to quickly identify new outbreaks and track their spread. This includes the use of tools such as web-based reporting systems, mobile health (mHealth) applications, and social media monitoring.
      \item \textbf{\textit{Contact Tracing and Case Management:}} Contact tracing is a critical component of pandemic management, and it can be greatly facilitated through the use of ICT. This includes digital contact tracing apps, case management systems for healthcare providers, and digital tools for self-reporting and self-isolation monitoring.
      \item \textbf{\textit{Data Analytics and Visualization:}} Data analytics and visualization tools are essential for understanding the spread of the pandemic, identifying trends, and making informed decisions. This includes the use of dashboards, data visualizations, and predictive analytics.
      \item \textbf{\textit{Communication and Education:}} Effective communication is crucial during a pandemic, and ICT can be used to disseminate information to the public, healthcare providers, and other stakeholders. This includes the use of digital communication channels such as social media, chatbots, and online forums, as well as digital education and training resources.
      \item \textbf{\textit{Telemedicine and Remote Care:}} The use of telemedicine and remote care can help to reduce the burden on healthcare systems, minimize exposure to the virus, and improve access to care. This includes the use of teleconsultation platforms, remote monitoring tools, and digital health records.
\end{enumerate}
 The transitioning of these key functionalities towards digitization, automation and intelligences are crucial to improve patient care quality and capacity during outbreaks; which were severely impacted during the pandemic. Given the current shortcomings in healthcare systems, steps must be taken to modernise and adapt healthcare systems in order to better prepare for future pandemics.
 As the world continues to navigate the COVID-19 pandemic, it has become clear that healthcare ICT systems will play an increasingly important role in the future of healthcare delivery. In a post COVID-19 world, there are several ICT strategies that should be implemented to improve the quality, efficiency, and accessibility of healthcare services. These include \cite{Filip2022}:
\begin{enumerate}
    \item \textbf{\textit{Expansion of Telemedicine:}} Telemedicine has become an essential tool during the pandemic for remote consultations and monitoring of patients. In a post-COVID-19 world, healthcare providers should continue to expand telemedicine services to provide remote care and support to patients in underserved areas, as well as those who are unable to travel to healthcare facilities.
    \item \textbf{\textit{Data Analytics and Population Health Management:}} Healthcare ICT systems should prioritize data analytics and population health management to identify high-risk populations and provide targeted interventions. This includes the use of predictive analytics to identify patients who are at risk of developing certain conditions, as well as the implementation of population health management tools to support chronic disease management and preventive care.
    \item \textbf{\textit{Remote Patient Monitoring:}} Remote patient monitoring tools can help healthcare providers to manage chronic conditions and monitor patients remotely. In a post-COVID-19 world, healthcare providers should expand the use of remote patient monitoring tools to reduce the need for in-person appointments and support patients in managing their health outside of the clinical setting.
    \item \textbf{\textit{Interoperability:}} Interoperability is key to improving the efficiency and effectiveness of healthcare delivery. In a post COVID-19 world, healthcare ICT systems should prioritize the development of interoperable systems to enable the seamless sharing of data between healthcare providers, patients, and different healthcare systems.
    \item \textbf{\textit{Cybersecurity:}} With the increased use of healthcare ICT systems, cybersecurity will become even more critical. Healthcare organizations should prioritize cybersecurity to ensure that patient data is protected and secure, and to prevent cyber attacks that could disrupt healthcare delivery.
\end{enumerate}

During the early stages of the pandemic, protocols for treating COVID-19 cases were frequently changing due to a lack of information, insufficient supplies and capabilities of healthcare facilities, limited availability of effective prophylactics, and disagreements over treatment methods. For example, the use of the antiparasitic drugs as a treatment for COVID-19 was widespread in various countries, but its effectiveness was not supported by clinical evidence \cite{Molento2021}, \cite{Hellwig2020}. Additionally, there were disruptions in the supply chain of certain drugs \cite{Hellwig2020}, leading to shortages and the use of alternative medications. These issues highlight the need for better preparedness in healthcare systems to manage pandemics and ensure access to necessary medical supplies and treatments, especially in areas of lower economic status. Effective planning is essential to protect against the spread of pathogens and minimize lockdowns and restrictions \cite{Bader2020}. Preventive measures are equality important with the establishment of streamlined processes for the development of vaccines and innovative treatments through understanding of molecular and genomic aspects  \cite{Ferrinho2020}. This process necessitates cross-boarder collaboration for spatio-temporal tracking of viruses across large geographical regions and constant assessment of genetic variations \cite{Huf2021}.

The ongoing impact of shutdowns on healthcare systems, particularly in underdeveloped countries, is still apparent. The last two years have shown the urgent need for reviving and restructuring medical services worldwide. This includes improving ICU triage, flexible increase in ICU capacity and staffing, ensuring safety in facility design, having enough supplies and equipment, focusing on the well-being of healthcare workers, and incorporating better end-of-life care in hospital management \cite{Pleyers2020}. The pandemic also highlighted the significance of digital transformation and streamlining of epidemiological registries in healthcare. Reorganising healthcare systems should be a priority to improve efficiency and availability of drugs, preventive medicine, tele-medicine, and reduce emergency and hospital visits. The SARS-CoV-2 pandemic has demonstrated that this is crucial for global health, regardless of location \cite{Baks2021}, \cite{Dyal2020}, \cite{Zeina2021}.

It also becomes increasingly clear that the development and distribution of vaccines worldwide was never enough to end the COVID-19 pandemic, even though considerable resources have been invested in design, production, and distribution of vaccines \cite{Swan2021}. Evidently, people in countries with high vaccination rates have less severe symptoms and reduced hospitalisation rates, but vaccines do not prevent infection and transmission \cite{Zeina2021}. Effective control of the virus will depend on the creation of vaccination platforms with priority groups, deployment of vaccination centres, and trust in medical professionals. Transparency of government authorities, research experts, and healthcare professionals is also crucial to inform the public about the safety and side effects of vaccines compared to the risk of COVID-19 \cite{Dascalu2021}. To make informed decisions, a system to report potential side effects and other issues concerning vaccination is necessary. Despite these challenges, the future of healthcare has gone online, with many advantages for patients and healthcare providers.

Overall, this survey aims to provide a holistic review of the key components around public health surveillance systems and the necessary transformations that need to take place in order to be better prepared for future incidents and emergencies. The survey is organised as follows. Section \ref{HealthcareICTSystems} describes three categories of Healthcare ICT systems, the Electronic Health Records (EHR), Hospital Management Systems (HMS) and Health Tracking \& Monitoring Systems and their challenges. Section \ref{PublicHealthcareNeeds} provides the Public Healthcare needs and its challenges as experienced during the COVID-19 pandemic. Specifically, we emphasised on Epidemiological Surveillance, Case Identification and Public Communication Platforms. Section \ref{PublicHealthcareEmergency} reveals the emergency response ICT challenges in the public Health sector during COVID-19. Section \ref{Discussion} discusses key findings and highlights important outcomes and future developments. Section \ref{Conclusions} provides concluding remarks and highlights a number of important lessons learned during the COVID-19 pandemic in order to be better prepared to future pandemics.

\section {Healthcare ICT systems}
\label{HealthcareICTSystems}
%just refer the 11 categories, and why we keep only below three

Healthcare may be one of the most complex industries to digitalize as it involves many stakeholders: from hospitals and medical centers, through doctors, pharmaceutical and research companies, up to patients themselves. Moreover, Healthcare Information and Communication Technology (ICT) systems refer to the use of technology in healthcare to manage and process information, as well as to improve communication and collaboration among healthcare professionals \cite{ariani2017innovative}. These systems can include electronic health records (EHRs), telemedicine, mHealth, and health information exchange (HIE). One of the main benefits of healthcare ICT systems is the ability to improve the quality and safety of care \cite{nyame2020improving}. Electronic health records (EHRs), for example, allow for the storage and sharing of patient information in a secure and efficient manner, which can improve the accuracy of diagnoses and treatment plans. Telemedicine and mHealth can also improve access to care, particularly for those in remote or underserved areas \cite{nyame2020improving}. Another benefit of healthcare ICT systems is the ability to improve efficiency and reduce costs. Electronic health records (EHRs) can reduce the need for paper records and reduce administrative tasks, while telemedicine and mHealth can improve the coordination of care and reduce the need for in-person visits. Health information exchange (HIE) allows for the sharing of patient information across different healthcare organizations, which can improve continuity of care, reduce errors and support population health management. One of the main challenges of healthcare ICT systems is the need for robust and secure systems to protect patient data. This includes issues such as data privacy and security, as well as the need for systems that are reliable and can withstand cyber attacks. Another challenge is the need for interoperability, which means that different systems need to be able to communicate and share information with each other in a seamless and efficient manner. Finally, there is a need to ensure that healthcare ICT systems are user-friendly and easy to use, particularly for healthcare professionals who may not have a lot of experience with technology \cite{colombo2020health}. In conclusion, healthcare ICT systems are an important aspect of modern healthcare. They can improve the quality and safety of care, increase efficiency, and reduce costs. However, it is important to address challenges such as data security, interoperability, and user-friendliness in order to ensure that these systems are effective and beneficial for both patients and healthcare professionals. Below we describe three categories that are more relevant with the subject matter of this survey.

\subsection{Electronic Health Records (EHR)}
%%Done
The electronic health record (EHR) has been gaining popularity as a source of patient data for both clinical and research purposes \cite{Binkeheder2021}. EHRs allow for the digitalization and centralization of patient health information, making it more accessible and usable for healthcare providers, which can improve continuity of care and coordination between different providers, as well as enable population health management and analytics \cite{Xu2015}. EHRs were initially designed for clinical care and billing purposes \cite{Denny2012}. However, there has been an effort to optimize the types of data and methods of storing this data to support research initiatives. The primary types of information available from EHRs include billing data, laboratory results and vital signs, provider documentation, documentation from reports and tests, and medication records \cite{Denny2012}. This data can be used for a variety of secondary purposes \cite{Binkeheder2021} such as decision-making, managing patient conditions, data exchange, building predictive models and discovering new medical knowledge.

EHRs are widely used in epidemiological research, but the validity of the results is dependent upon the assumptions made about the healthcare system, the patient, and the provider \cite{gianfrancesco2021narrative}. 
\begin{figure}[htbp]
    \centering
    \includegraphics[width=0.7\textwidth]{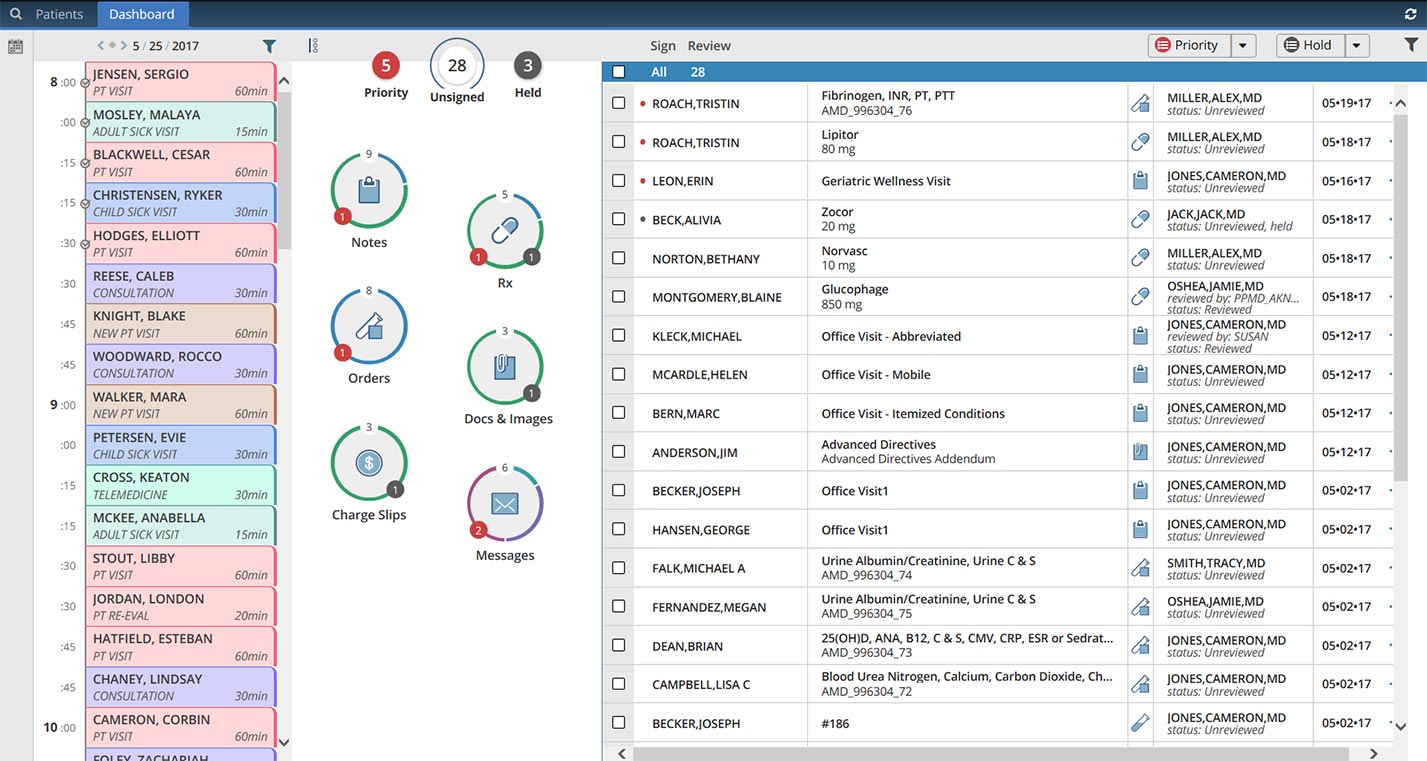}
    \caption{AdvancedMD EHR Software \cite{advancedmd_2022}}
    \label{fig:AdvancedEHRSoftware}
\end{figure}

\subsubsection{Data Available in EHRs}
\begin{itemize}

\item {\textbf{\textit{Billing data}}}: They typically consists of codes derived from the International Classification of Diseases (ICD) and Current Procedural Terminology (CPT). ICD is a terminology system for diseases, symptoms, and procedures that is maintained by the World Health Organization (WHO). It is widely used across the world and most countries use version 11 \cite{icd11who}. These codes are commonly used in Electronic Medical Record (EMR) systems and are also used for research purposes \cite{herzig2009acid,klompas2008automated,kiyota2004accuracy,dean2009use,elixhauser1998comorbidity,charlson1987new}, they are often used to bill insurance companies. However, previous research \cite{li2008comparing,elkin2001randomized} has shown that these codes can have low accuracy and specificity. Despite this, they are still used as part of more complex algorithms that have a high level of performance \cite{ritchie2010robust,liao2010electronic,conway2011analyzing}. CPT codes are created and maintained by the American Medical Association and are used for billing clinical services, they are often paired with ICD codes to satisfy the requirements of insurance companies. For example, insurance companies will pay for a brain MRI scan if it is ordered for certain symptoms or diseases but not for others \cite{faciszewski2003procedural}.

\item {\textbf{\textit{Laboratory and Vital Signs}}}: The medical record contains a long-term record of mostly structured data in the form of laboratory data and vital signs. These data can be stored as name-value pairs, and can also be encoded using standard terminologies. The most commonly used controlled vocabulary for laboratory tests and vital signs is the Logical Observation Identifiers Names and Codes (LOINC), which is a Consolidated Health Informatics standard for the representation of laboratory and test names and is part of Health Language 7 (HL7) \cite{huff1998development,hl7international}. Despite the increasing use of LOINC, many hospital lab systems still use local dictionaries to encode laboratory results internally. This can cause issues as hospital laboratory systems or testing companies may change over time, resulting in different internal codes for the same test result, so care must be taken to implement selection logic based on laboratory results \cite{sreenivasan2021interoperability}.

\item {\textbf{\textit{Provider Documentation}}}: Clinical documentation is a valuable source of information for identifying patients with specific characteristics or conditions, known as phenotyping. This documentation, which is required for billing and can be found in electronic health record systems, must be in an electronic format for it to be useful for phenotyping efforts \cite{rosenbloom2010generating}. This can include unstructured narrative text or structured documentation, which can be processed using text searches or natural language processing techniques.

\item {\textbf{\textit{Documentation from Reports and Tests}}}: The provider-generated reports and test results include things like radiology and pathology reports and some procedure results such as echocardiograms. They often come in the form of narrative text results \cite{denny2005identifying}, which can include both structured and unstructured data. For example, an electrocardiogram report may have structured data like interval durations, as well as a narrative text interpretation provided by the cardiologist. These structured data are generated by automated algorithms and may vary in accuracy \cite{willems1991diagnostic}.

\item {\textbf{\textit{Medication Records}}}: Medication records play a crucial role in identifying and characterizing a patient's condition. They can be used to improve the accuracy of identifying a particular case, and to confirm that patients believed to be healthy do not have the disease. Medications received by a patient serve as evidence that the treating physician believed that the disease was present and required treatment. Medication records can be in different forms within an electronic record. With the widespread use of computerized provider order entry systems in hospitals, inpatient medication records are often available in highly structured records that can be mapped to controlled vocabularies. Many hospitals are also using automated barcode medication administration \cite{poon2010effect} records, which allow for accurate drug exposure and timing to be recorded for each inpatient. Even without these systems, research has shown that medications ordered through CPOE systems are administered with high reliability \cite{fitzhenry2007medication}.
\end{itemize}

In early 90s EHR systems were mostly developed and used at academic centers as a hybrid of paper and electronic data. These systems are based on hierarchical and relational databases and are mainly focused on billing and scheduling systems, as well as some clinical systems. They were mostly running on large mainframe and minicomputers with limited data storage capacity, but with the advancement of technology, personal computers with graphics were used as monitors. Data entry was mainly done through keyboards and mouse. EHR systems were used in both inpatient and outpatient facilities and were connected through Local area networks and the Internet, with many systems being web-based. They included features such as image scanning and paper printouts, Clinical Decision Support, Computerized Provider Order Entry, drug references, clinical manuals, textbooks of medicine, literature searching, physician documentation, and electronic signatures. They also complied with Health Level 7 and IEEE P1157 standards and Universal Medical Language System. There were also interfaces for medical devices and Picture Archiving and Communication Systems (PACS) were used to store and retrieve images. However, there were also ethical issues that need to be considered such as data ownership, data liability, informed consent, security, and privacy.

In 2022 EHR systems are widely used in health care, nursing homes, insurance companies, and relevant governmental departments. They are mostly vendor EHRs which are interfaced with Personal Health Records. Intra-facility data sharing with standards is becoming more common and data mapping to SMOMED \& LOINC \cite{bodenreider2018recent} is also used. They are also used in digital pathology, mental health, external labs, e-prescribing, order-sets, family history, genetics, biobanks, biosurveillance, and public health. With the increase of EHRs, Clinical Decision Support has also increased and it is now incorporating Natural Language Processing and Big Data. They are also being used in mobile devices and open-source EHRs are also becoming more prevalent. The use of digital identities and enterprise data warehouses for example to monitor birth to death information, are also increasingly supported by theses systems. Cloud computing is also being used more often and notifications/alerts are increasingly prevailing (for instance via email, pagers and cell phone sms) to inform relevant users of changes. However, there are also ongoing challenges for introducing further information and new features that have significantly increased interest in EHR related publications.

\subsection{Hospital Management Systems (HMS)}
%%Done
A hospital management system (HMS) \cite{vegoda1987introduction} is a computerized system that is used to improve patient care by increasing knowledge and reducing uncertainty. It allows for rational decisions to be made based on the most recent information provided. The system streamlines operations, improves administration and control, and enhances patient care while also controlling costs and improving profitability. HMS is designed to be powerful, flexible, and easy to use \cite{berg1999patient, kuhn2001hospital}. It is made up of various subsystems that are integrated to capture data in different sections of a hospital, including patient registration, billing, laboratory, pharmacy, and radiology. These subsystems include \cite{gupta2020hospital, garrido2004making}:
\begin{figure}[htbp]
    \centering
    \includegraphics[width=0.7\textwidth]{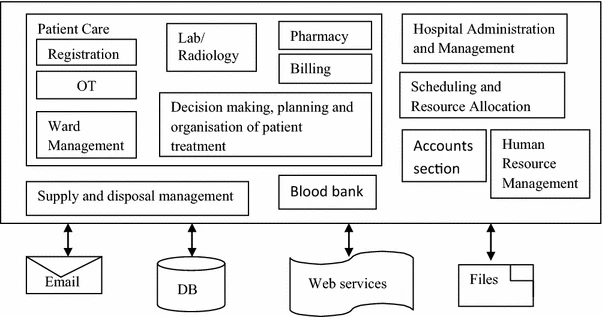}
    \caption{HMS High Level Architecture \cite{angeline2018case}}
    \label{fig:HMSHighLevelArchitecture}
\end{figure}

\begin{itemize}
\item{\textbf{\textit{Appointment Management}}}\cite{gupta2008appointment}: Hospital appointment management platforms (see Figure \ref{fig:HosAppManagementPlatArch}) are digital tools that help hospitals and medical clinics to streamline their patient scheduling and appointment booking processes. They provide a centralized system that allows patients to easily book appointments online, while also giving healthcare providers access to real-time patient schedules and appointment information. Some of the key features of these platforms include:
\begin{enumerate}
    \item \textbf{\textit{Online appointment booking}}: Patients can easily book appointments online, choosing a date and time that works for them.
    \item \textbf{\textit{Real-time schedules}}: Healthcare providers can view real-time schedules and appointment information, making it easy to manage patient flow and avoid overbooking.
    \item \textbf{\textit{Automated reminders}}: Patients can be automatically reminded of upcoming appointments via text or email, reducing the risk of missed appointments.
    \item \textbf{\textit{Patient self-scheduling}}: Patients can schedule their own appointments and view available slots in real-time, reducing the need for staff to manage the process.
    \item \textbf{\textit{Integrations}}: These platforms can be integrated with electronic health records (EHRs) or other hospital systems, allowing for seamless transfer of patient data.
    \item \textbf{\textit{Reporting}}: Platforms can generate reports on patient scheduling and appointment information, allowing hospitals to analyze data and improve patient flow.
\end{enumerate}
Overall, hospital appointment management platforms can help to improve patient satisfaction and reduce administrative burdens on staff, by allowing patients to easily schedule appointments, and providing real-time scheduling data to healthcare providers.

\begin{figure}[htbp]
    \centering
    \includegraphics[width=0.7\textwidth]{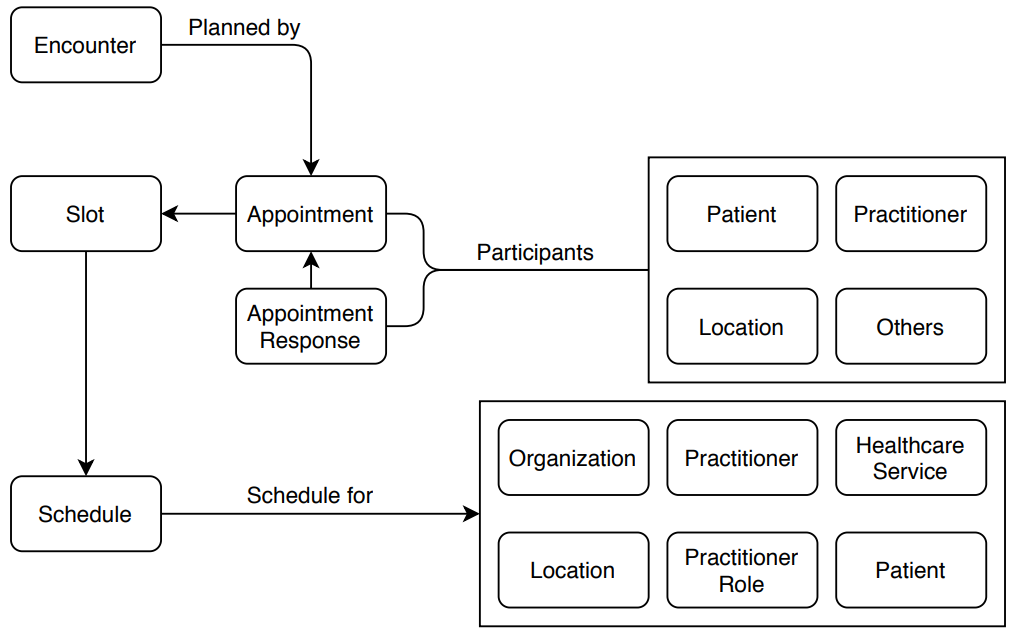}
    \caption{Hospital Appointment Management Platform Architecture \cite{chaves2021development}}
    \label{fig:HosAppManagementPlatArch}
\end{figure}

\item{\textbf{\textit{Patient Management}}}\cite{douglas2003lilongwe}: Hospital Patient Management Platforms (HPMP) are computer-based systems used by healthcare organizations to manage and track patient information. These systems typically include features such as:
\begin{enumerate}
\item \textbf{\textit{Patient registration and demographic information management}}: This feature allows for the capturing and storage of patient personal and contact information, insurance details, and other demographic data.
\item \textbf{\textit{Appointment scheduling and management}}: By using this function, patients you will be able to schedule and manage appointments online, view upcoming appointments, and receive reminders. It also enables the healthcare provider to manage their schedules and ensure the availability of resources.
\item \textbf{\textit{Electronic Medical Records (EMR) and charting}}: With the implementation of this tool, healthcare providers can document and access patient medical history, treatments, and lab results electronically. It also enables the sharing of patient information among different healthcare providers.
\item \textbf{\textit{Clinical decision support and treatment planning}}: This new functionality offers healthcare providers the opportunity to access patient data, guidelines and best practices to make informed decisions about patient care, and develop personalized treatment plans.
\item \textbf{\textit{Medication management and prescription writing}}: Healthcare providers are now able to electronically prescribe medications, track dosages and refill requests, and alert providers to potential drug interactions.
\item \textbf{\textit{Laboratory and diagnostic test ordering and results management}}: This tools enables healthcare providers to electronically order and review laboratory and diagnostic test results.
\item \textbf{\textit{Patient billing and insurance management}}: This tool can easily allows the management of patient billing, insurance claims, and payments. It also enables the tracking of outstanding balances and insurance reimbursements.
\item \textbf{\textit{Patient education and self-management tools}}: This feature ensures that the patients with educational resources, self-management tools, and access to their medical information to improve their understanding of their health conditions and treatment options.
\item \textbf{\textit{Care coordination and communication with other healthcare providers}}: By utilizing this feature, healthcare providers can easily communicate and collaborate with other providers, such as specialists and primary care physicians, to coordinate patient care.
\item \textbf{\textit{Population health management and analytics}}:By utilizing this feature, healthcare providers can easily to track and manage patient populations, identify trends and patterns, and monitor the effectiveness of treatment plans.
\item \textbf{\textit{Secure messaging and telemedicine capabilities}}: This feature allows patients and providers to communicate securely via messaging and videoconferencing, enabling remote consultations and follow-up.
\item \textbf{\textit{Reporting and data analytics for patient outcomes and performance metrics}}: The introduction of this feature means that healthcare providers can now access patient data and analytics to track outcomes, monitor performance, and identify areas for improvement.
\item \textbf{\textit{Mobile access for patients and providers}}: This tool enables patients and providers to access the system and patient data on their mobile devices.
\item \textbf{\textit{Compliance with regulatory standards such as HIPAA and Meaningful Use}}: This functionality ensures that the system is compliant with regulations such as the Health Insurance Portability and Accountability Act (HIPAA) and the Meaningful Use (MU) program.
\end{enumerate}

Overall, HPMPs are computer-based systems that allow healthcare organizations to manage and track patient information, including electronic health records, appointment scheduling, medication management, lab results tracking, analytics, and reporting. The goal of HPMP is to improve the quality of care for patients by providing healthcare providers with quick and easy access to patient information, which can help to reduce errors, improve communication, and make data-driven decisions.
\begin{figure}[htbp]
    \centering
    \includegraphics[width=0.7\textwidth]{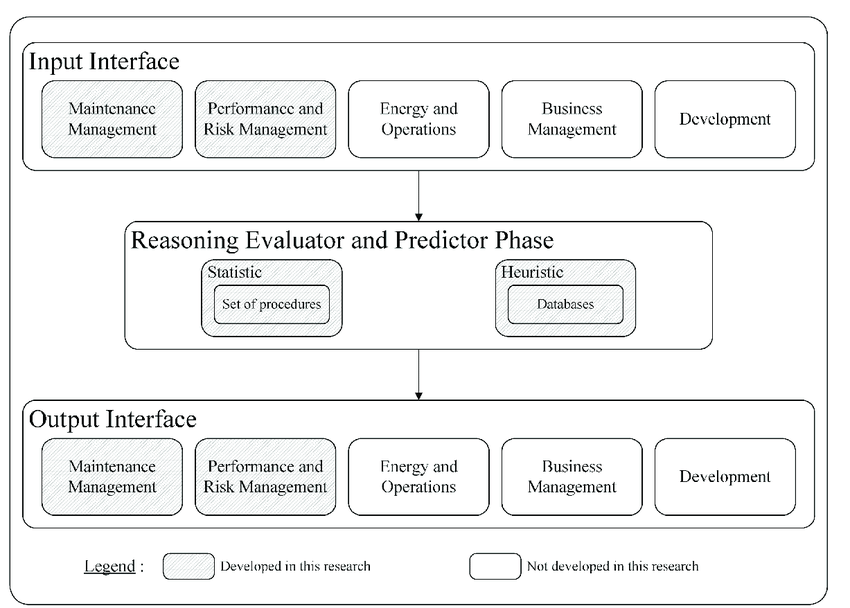}
    \caption{Healthcare Facility Management Model \cite{lavy2007integrated}}
    \label{fig:HealthcareFacilityManagement}
\end{figure}

\item{\textbf{\textit{Facility Management}}}\cite{lai2019performance}: Hospital facility management platforms are digital tools that help hospitals and medical clinics to manage their physical facilities and infrastructure. They provide a centralized system for tracking and monitoring various aspects of a hospital's physical operations, such as maintenance, safety, and energy efficiency. Some of the key features of these platforms include:
\begin{enumerate}
\item \textbf{\textit{Room Management}}: Provides a comprehensive solution for managing the allocation and availability of rooms, as well as scheduling regular maintenance tasks.
\item \textbf{\textit{Equipment Management}}: Enables tracking and maintenance of medical equipment, ensuring that all equipment is functioning properly, and maintenance tasks are scheduled as needed.
\item \textbf{\textit{Inventory Management}}: Medical supplies can be tracked, reordered, and usage can be monitored to ensure that inventory levels are optimized.
\item \textbf{\textit{Maintenance Management}}: Facility managers can schedule and track maintenance tasks, ensuring that all systems are functioning optimally and that any issues are addressed in a timely manner.
\item \textbf{\textit{Energy Management}}: Enables tracking and management of energy usage, which can lead to cost savings and more efficient energy use.
\item \textbf{\textit{Security Management}}: Provides tools for monitoring and managing security systems, including access control and surveillance cameras, ensuring that the facility is secure at all times.
\item \textbf{\textit{Asset Management}}: All assets within the facility, including furniture and equipment, can be tracked and managed, ensuring that they are properly maintained and accounted for.
\item \textbf{\textit{Compliance Management}}: Enables monitoring and management of facility compliance with local and national regulations, including HIPAA and OSHA compliance, ensuring that the facility is operating in a safe and compliant manner.
\item \textbf{\textit{Reporting and Analytics}}: Provides reporting and analytics tools for facility usage, maintenance, and energy usage, enabling data-driven decision-making and more effective management.
\item \textbf{\textit{Mobile Access}}: Remote monitoring and management of the facility through mobile devices it is allowed by this platform, ensuring that managers can stay up-to-date on facility operations even when they are not on-site.
\end{enumerate}
Overall, hospital facility management platforms can help to improve the efficiency and effectiveness of a hospital's physical operations, by providing a centralized system for tracking and monitoring various aspects of facility management and by providing real-time visibility into the hospital's physical operations.

\item{\textbf{\textit{Staff Management}}}\cite{lakbala2013knowledge}: Hospital staff management platforms are digital tools that help hospitals and medical clinics to manage and organize their staff. They provide a centralized system for tracking and monitoring staff schedules, time-off, and performance, as well as providing tools for communication and collaboration among staff members. Some of the key features of these platforms include:
\begin{enumerate}
\item \textbf{\textit{Employee Management}}: Facilitates the tracking and management of employee information, including personal details, job titles, and schedules.
\item \textbf{\textit{Shift Management}}: By using shift management tools, managers can schedule and manage employee shifts, allowing for shift swapping and effective overtime management.
\item \textbf{\textit{Attendance Management}}: Attendance management software enables the tracking and management of employee attendance, including leave requests and time-off management.
\item \textbf{\textit{Payroll Management}}: With payroll management software, managers can efficiently handle employee payroll, including calculating pay, taxes, and deductions, and generating paystubs and reports.
\item \textbf{\textit{Performance Management}}: Performance management tools enable managers to track and manage employee performance, including setting goals, tracking progress, and providing feedback.
\item \textbf{\textit{Training Management}}: Training management software allows managers to track and manage employee training, including scheduling and tracking progress.
\item \textbf{\textit{Recruitment Management}}: Recruitment management software streamlines recruitment processes by enabling posting of job openings, tracking resumes, and scheduling interviews.
\item \textbf{\textit{Reporting and Analytics}}: Reporting and analytics software generates insightful reports and analytics on employee data, including attendance, performance, and turnover rate.
\item \textbf{\textit{Mobile Access}}: Mobile access to employee management platforms allows managers to remotely monitor and manage staff using their mobile devices.
\item \textbf{\textit{Communication}}: By facilitating internal communication through messaging, discussion boards, and shared documents, communication tools on employee management platforms foster collaboration and engagement among staff.
\end{enumerate}
Overall, hospital staff management platforms can help to improve the efficiency and effectiveness of a hospital's staff, by providing a centralized system for tracking and monitoring staff schedules, time-off, and performance, and by providing tools for communication and collaboration among staff members.

\begin{figure}[htbp]
    \centering
    \includegraphics[width=0.7\textwidth]{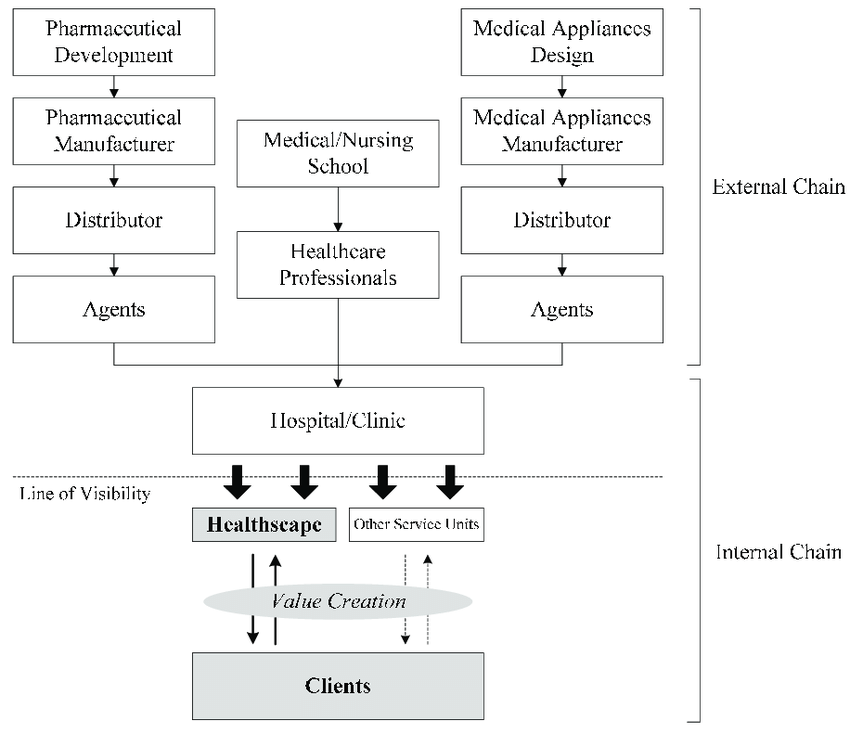}
    \caption{Healthcare Supply Chain Structure \cite{hsu2019facilitating}}
    \label{fig:HealthcareSupplyChainStructure}
\end{figure}

\item{\textbf{\textit{Supply Management}}}\cite{landry2013challenges}: Hospital supply management platforms are digital tools that help hospitals and medical clinics to manage their inventory of medical supplies, equipment and drugs. They provide a centralized system for tracking and monitoring the inventory levels, ordering, receiving, and distribution of these items. Some of the key features of these platforms include:
\begin{enumerate}
    \item \textbf{\textit{Inventory management}}: Offers a comprehensive solution for tracking inventory levels, managing stock levels, and preventing stockouts.
    \item \textbf{\textit{Ordering and receiving}} Hospitals can utilize the tools provided by these platforms to create purchase orders, monitor deliveries, and receive items into inventory efficiently.
    \item \textbf{\textit{Distribution and allocation}}: The distribution and allocation of medical supplies, equipment, and drugs to different departments and clinics within the hospital can managed by this platforms.
    \item \textbf{\textit{Automated reordering}}: By automatically reordering items when stock levels drop below a certain threshold, inventory management platforms minimize the risk of stockouts and help hospitals maintain optimal inventory levels.
    \item \textbf{\textit{Reporting}}: Inventory management platforms generate a variety of reports on inventory levels, costs, usage, and other metrics, enabling managers to identify trends and make informed decisions.
    \item \textbf{\textit{Supply Chain Visibility}}: Supply chain visibility is greatly enhanced by inventory management platforms, which provide real-time visibility of the entire supply chain process, from order placement to delivery.
    \item \textbf{\textit{Electronic Data Interchange (EDI)}}: By leveraging electronic data interchange (EDI), inventory management platforms streamline the purchase order process and ensure timely delivery of items by communicating seamlessly with suppliers, distributors, and other partners.
\end{enumerate}
Overall, hospital supply management platforms can help to improve the efficiency and effectiveness of a hospital's supply chain by providing a centralized system for tracking and monitoring inventory levels, ordering, receiving, and distribution of medical supplies, equipment and drugs, and providing real-time visibility of the entire supply chain process.
\item{\textbf{\textit{Financial Management}}}\cite{srinivasan2008managing}: Hospital financial management platforms are digital tools that help hospitals and medical clinics to manage their financial operations. They provide a centralized system for tracking and monitoring revenue, expenses, billing, and other financial activities. Some of the key features of these platforms include:
\begin{enumerate}
\item \textbf{\textit{Billing and Invoicing}}: Generation and management of patient bills, including insurance claims and payment processing, are enabled by the Billing and Invoicing feature
\item \textbf{\textit{Accounts Payable}}: Vendor invoices, purchase orders, and payments can be managed with ease thanks to the Accounts Payable feature.
\item \textbf{\textit{Accounts Receivable}}: The management of patient payments, including patient statements and payment collections, is made possible by the Accounts Receivable feature.
\item \textbf{\textit{Revenue Cycle Management}}: The Revenue Cycle Management feature provides a comprehensive solution for managing the entire revenue cycle, from patient registration to insurance claims and payments.
\item \textbf{\textit{Budgeting and Forecasting}}: Creation and management of budgets, including forecasting future revenue and expenses, is made possible by the Budgeting and Forecasting feature.
\item \textbf{\textit{General Ledger}}: Financial transactions, including journal entries and account balances, can be managed easily using the General Ledger feature.
\item \textbf{\textit{Financial Reporting}}: Financial Reporting feature allows for the generation of financial reports, including income statements, balance sheets, and cash flow statements, providing valuable insights into the hospital's financial health.
\item \textbf{\textit{Cost Analysis}}: The Cost Analysis feature enables analysis of costs, including resource utilization, supply chain, and labor costs, helping hospitals to identify areas for improvement and cost savings.
\item \textbf{\textit{Auditing and Compliance}}: The Auditing and Compliance feature provides a tool for managing audit and compliance processes, including HIPAA, CMS, and other regulatory requirements, ensuring that hospitals comply with all necessary regulations.
\item \textbf{\textit{Integration with other systems}}: The integration with other hospital systems, such as EHRs and supply chain management systems, to streamline financial processes and ensure accurate data is utilised with this feature.
\end{enumerate}
Overall, hospital financial management platforms can help to improve the efficiency and effectiveness of a hospital's financial operations by providing a centralized system for tracking and monitoring revenue, expenses, billing, and other financial activities. They can also provide tools for budgeting, forecasting, and cost accounting, which can help managers to make data-driven decisions and optimize the financial performance of the hospital.
\item{\textbf{\textit{Insurance Management}}}\cite{falah2022characterisation}: Hospital insurance management platforms are digital tools that help hospitals and medical clinics to manage their interactions with insurance providers and payers. They provide a centralized system for tracking and monitoring insurance claims, payments, and other insurance-related activities. Some of the key features of these platforms include:
\begin{enumerate}
\item \textbf{\textit{Claims management}}: process and manage insurance claims, including submission, tracking, and approval processes.
\item \textbf{\textit{Patient enrollment and eligibility verification}}: verify patient insurance coverage and enrollment status.
\item \textbf{\textit{Billing and revenue cycle management}}: generate and submit bills to insurance companies, track payments and denials, and manage the revenue cycle.
\item \textbf{\textit{Contract management}}: manage contracts with insurance companies, including rate negotiations and compliance with contract terms.
\item \textbf{\textit{Fraud detection and prevention}}: identify and prevent fraudulent claims and billing practices.
\item \textbf{\textit{Reporting and analytics}}: generate reports and analyze data on insurance claims, payments, denials, and other key metrics.
\item \textbf{\textit{Integration with other systems}}: integrate with other hospital systems, such as electronic health records and scheduling systems, to streamline processes and improve data accuracy.
\item \textbf{\textit{Policy management}}: manage and track insurance policies and ensure compliance with regulations.
\item \textbf{\textit{Communication with insurance companies}}: communicate with insurance companies via electronic data interchange (EDI) or other methods to ensure timely and accurate claim processing.
\item \textbf{\textit{Security and compliance}}: ensure compliance with regulatory requirements for patient data protection and security.
\end{enumerate}
Overall, hospital insurance management platforms can help to improve the efficiency and effectiveness of a hospital's interactions with insurance providers and payers. They provide a centralized system for tracking and monitoring insurance claims, payments, and other insurance-related activities, and can help to ensure that hospitals and clinics receive the reimbursement they are entitled to. The platform's functionality can also help to automate the process and reduce administrative burdens on staff.
\item{\textbf{\textit{Laboratory Management}}}\cite{allen2013role}: Hospital laboratory management platforms are digital tools that help hospitals and medical clinics to manage their laboratory operations. They provide a centralized system for tracking and monitoring lab results, test ordering, and other laboratory-related activities. Some of the key features of these platforms include:
\begin{enumerate}
\item \textbf{\textit{Patient Information Management}}: The storage and retrieval of patient data, such as demographics, medical history, and test results, is made possible by the Managing Patient Information feature.
\item \textbf{\textit{Test Ordering and Result Reporting}}: This feature is a tool that enables tracking and management of laboratory test orders, as well as the generation and distribution of test results.
\item \textbf{\textit{Inventory Management}}: With the Inventory Control feature, laboratory supplies and reagents can be managed and tracked more efficiently.
\item \textbf{\textit{Billing and Insurance}}: Facilitates the process of billing patients and insurance companies for laboratory services.
\item \textbf{\textit{Electronic Medical Records (EMR) Integration}}: This tool is responsible for tracking and monitoring quality control measures and compliance with regulatory standards, ensuring a high level of quality in laboratory testing.
\item \textbf{\textit{Quality Control and Compliance}}: For tracking and monitoring quality control measures and compliance with regulatory standards, ensuring a high level of quality in laboratory testing.
\item \textbf{\textit{Reporting and Analytics}}: Generates various reports and analytics, including patient statistics and test volume, providing valuable insights into laboratory performance.
\item \textbf{\textit{Security and Access Control}}: Ensures secure access to patient information and laboratory data, with roles-based access control.
\item \textbf{\textit{Mobile and Remote Access}}: Allows laboratory information to be accessed remotely from different locations and devices, providing greater flexibility and convenience.
\item \textbf{\textit{Cloud-based Deployment}}: The deployment of the platform is allowed in the cloud, which enables scalability and easy access to the system from anywhere.
\end{enumerate}
Overall, hospital laboratory management platforms can help to improve the efficiency and effectiveness of a hospital's laboratory operations by providing a centralized system for tracking and monitoring lab results, test ordering, and other laboratory-related activities. The platform's functionality can also help to automate processes, ensure compliance and improve the quality of lab results and patient care.
\item{\textbf{\textit{Radiology Management}}}\cite{halsted2008design}: Hospital radiology management platforms are digital tools that help hospitals and medical clinics to manage their radiology operations. They provide a centralized system for tracking and monitoring radiology exams, images, and other radiology-related activities. Some of the key features of these platforms include:
\begin{enumerate}
\item \textbf{\textit{Radiology imaging management}}: Allows for the storage, retrieval, and viewing of radiology images such as X-rays, CT scans, and MRI images.
\item \textbf{\textit{Order management}}: Allows for the creation and tracking of radiology orders and exams, including scheduling, patient information, and exam results.
\item \textbf{\textit{Reporting and analysis}}: Generates reports and provides tools for analyzing radiology data, including patient statistics, exam trends, and equipment usage.
\item \textbf{\textit{Electronic Health Records (EHR) integration}}: Integrates with the hospital's EHR system to share patient information and radiology images with other healthcare providers.
\item \textbf{\textit{Workflow management}}: Allows for the management of radiology workflow and tasks, including the assignment of tasks to radiologists and technologists, and the tracking of progress.
\item \textbf{\textit{Image annotation and interpretation}}: Allows radiologists to annotate and interpret images, providing detailed information about the patient's condition.
\item \textbf{\textit{Quality assurance and compliance}}: Provides tools for ensuring compliance with regulatory standards, such as the HIPAA and DICOM, and for quality control of radiology images and reports.
\item \textbf{\textit{Advanced visualization tools}}: Includes advanced visualization tools such as 3D imaging, multi-planar reconstruction, and image fusion.
\item \textbf{\textit{PACS (Picture Archiving and Communication Systems) integration}}: Integrates with the hospital's PACS system to enable the storage, retrieval, and distribution of radiology images and reports.
\item \textbf{\textit{Remote access and collaboration}}: Allows remote access to radiology images and reports, and enables collaboration between radiologists and other healthcare providers.
\end{enumerate}
Overall, hospital radiology management platforms can help to improve the efficiency and effectiveness of a hospital's radiology operations by providing a centralized system for tracking and monitoring radiology exams, images, and other radiology-related activities. The platform's functionality can also help to automate processes, ensure compliance, improve patient care and streamline the radiology workflow.
\item{\textbf{\textit{Report Management}}}\cite{groene2013systematic}: Hospital report management platforms are digital tools that help hospitals and medical clinics to manage their medical reports. They provide a centralized system for generating, storing, and sharing medical reports, such as radiology, lab, and pathology reports. Some of the key features of these platforms include:
\begin{enumerate}
    \item \textbf{\textit{Report generation}}: The ability to generate and format medical reports, such as radiology, lab, and pathology reports, is provided by these platforms.
    \item \textbf{\textit{Report storage and sharing}}: These platforms offer tools for storing and sharing medical reports, such as integration with electronic medical records (EMR), making it easy for authorized providers and patients to access reports.
    \item \textbf{\textit{Report tracking}}: Tools for monitoring the status of medical reports, including report generation, review, and approval, are provided by these platforms.
    \item \textbf{\textit{Compliance}}: Can help hospitals and clinics to comply with regulatory requirements, such as HIPAA and DICOM, by providing tools for tracking and monitoring report management operations.
    \item \textbf{\textit{Reporting}}: Provides various reports on medical reports, such as report generation, review, and approval, which can help managers to identify trends and make data-driven decisions.
    \item \textbf{\textit{Automation}}: Automates many report management-related processes, such as report generation and sharing, to reduce administrative burdens on staff.
\end{enumerate}
Overall, hospital report management platforms can help to improve the efficiency and effectiveness of a hospital's report management operations by providing a centralized system for generating, storing, and sharing medical reports. The platform's functionality can also help to automate processes, ensure compliance, and improve patient care by providing easy access to medical reports for authorized providers and patients.
\item{\textbf{\textit{Support Management}}}\cite{allsop1995dealing}: Hospital support management platforms are digital tools that help hospitals and medical clinics to manage their administrative and support operations. They provide a centralized system for managing and organizing non-clinical activities such as human resources, payroll, and procurement. Some of the key features of these platforms include:
\begin{enumerate}
    \item \textbf{\textit{Human resources management}}: Provides tools for managing human resources activities such as employee records, payroll, benefits, and compliance with labor laws.
    \item \textbf{\textit{Payroll management}}: This is for managing payroll processes, including calculating, tracking and distributing employee salaries and benefits.
    \item \textbf{\textit{Procurement management}}: Managing the procurement process, including creating purchase orders, tracking deliveries and managing inventory.
    \item \textbf{\textit{Facility management}}: Platforms can provide tools for managing the hospital's facility, such as tracking and scheduling maintenance, monitoring energy consumption and compliance with safety regulations.
    \item \textbf{\textit{Reporting}}: Elavorates various reports on support operations, such as payroll, procurement, and facility management, which can help managers to identify trends and make data-driven decisions.
    \item \textbf{\textit{Automation}}: Automates many administrative and support-related processes, such as payroll and procurement, to reduce administrative burdens on staff.
\end{enumerate}
Overall, hospital support management platforms can help to improve the efficiency and effectiveness of a hospital's administrative and support operations by providing a centralized system for managing and organizing non-clinical activities. The platform's functionality can also help to automate processes, ensure compliance with regulations, and improve overall hospital operations.
\end{itemize}

To achieve these services it is mandatory to have high quality data storage, data speed, data exchange and networking. This is due to the variety as well as the amount of data that is collected and retrieved \cite{zachman1987framework, ferrara1997healthcare}. 

In order to model, validate and create HMS systems, several tools and technologies have been proposed through the years. For modelling, one of the dominant tools is Three-layer Graph based metal model ($3LGM^2$)\cite{winter2003modeling}, successor of $3LGM$\cite{winter1995three} which is a structured and validated approach \cite{schreiweis2010modelling}. In \cite{hubner2005specification} using $3LGM^2$ HMS approach was revised into a layer architecture.

In order to standardise HMS, American National Standards Institute (ANSI) introduced Health Level Seven (HL7) \cite{beeler1998hl7}. HL7 vision is to create an infrastructure for interoperability in the healthcare domain.

\subsection{Health Tracking \& Monitoring Systems} 
%%Done
Health tracking and monitoring is a crucial aspect of the healthcare industry as it allows medical doctors and researchers to gain insight into the health conditions of patients as well as the general population. Health Tracking \& Monitoring Systems (HTMS) play a vital role in proactively identifying and providing individuals with the assistance and services they need. This is made possible through the collection and analysis of contextual information. The success of these systems is dependent on their ability to collect and process data to understand a subject's environment and provide targeted services.

Contextual services in healthcare refer to a set of continuous processes that automatically acquire information about a subject, including behavioral, physiological, and environmental data. This information is then used to provide and adapt services accordingly, resulting in a better understanding of the subject's health conditions and improved care. This is achieved through the identification of behavior patterns and the ability to make more precise inferences about the subject's situation and environment. Numerous survey studies \cite{baig2013smart, chen2011body, peetoom2015literature, reeder2013framing} have been carried out in the literature, mainly focusing on smart home perspectives, health smart homes, wearable sensor-based systems, and healthcare applications for the elderly. These studies provide valuable information on the design and implementation of health tracking and monitoring systems, as well as the challenges and limitations of these systems.

Data collection is a crucial aspect of health tracking and monitoring systems, and the frequency of data collection plays an important role in the overall system performance. Data can be generated through three types of events: constant, instant, or interval. Constant events ensure continuous data transmission, instant events send data immediately upon the occurrence of a specific event, and interval events involve data being sent periodically at uniform time intervals. Physical sensors are commonly used to collect raw sensor data in HTMS, while virtual sources include existing health records and data from social media networks. These sensors are used to monitor various aspects of a subject's health and environment, including physiological data such as heart rate and blood pressure, environmental data such as temperature and humidity, and behavioral data such as activity levels and sleep patterns. The use of virtual sources can provide additional information on the subject's health history, behavior patterns, and social network, which can be used to improve the accuracy of the system's predictions and recommendations. Additionally, with the advent of IoT and connected devices, the use of wireless sensors and networked devices is becoming more prevalent in health tracking and monitoring systems. This allows for more accurate and real-time data collection, and enables remote monitoring of patients, which is particularly beneficial for elderly and disabled individuals who may have difficulty traveling to a healthcare facility for regular check-ups.
\begin{figure}[htbp]
    \centering
    \includegraphics[width=0.7\textwidth]{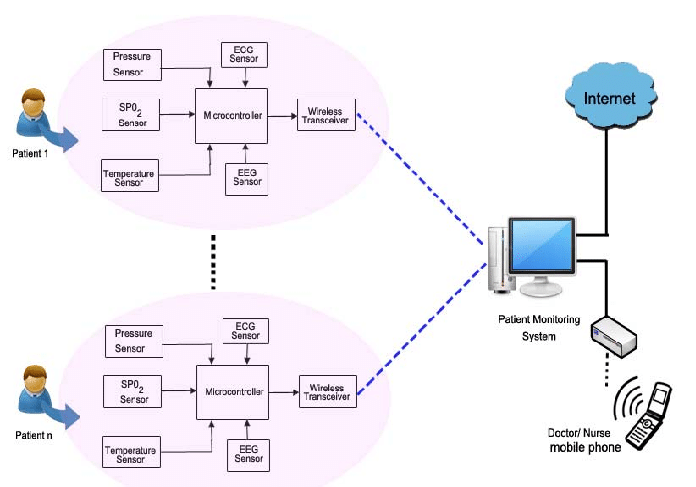}
    \caption{Patient Monitoring System Structure \cite{megalingam2012efficient}}
    \label{fig:PatientMonitoringSystemStructure}
\end{figure}

HTMS play a vital role in proactively identifying and providing individuals with the assistance and services they need. The success of these systems is dependent on their ability to collect and process data to understand a subject's environment and provide targeted services. The use of physical sensors and virtual sources, and wireless and networked devices, is becoming increasingly important in health tracking and monitoring systems, allowing for more accurate and real-time data collection, and enabling remote monitoring of patients.

Sensor data can come in various formats, such as numerical, categorical, graphics, video, etc. Based on these formats and sensor types, health monitoring can be divided into two approaches: vision-based and sensor-based approaches. Vision-based approaches rely on visual sensors, such as video cameras, for movement and gesture recognition, while sensor-based approaches use a wide range of emerging sensors and technologies for health and biomedical monitoring \cite{wood2008context, farella2010aware}. Recent systems and projects for healthcare monitoring commonly use standard or commercial sensors to gather raw data to monitor individuals and their environment. According to research projects surveyed in this paper, there are three main classes of interconnected networks that are often used in this field: Personal Sensor Networks (PSN), Body Sensor Networks (BSN), and Multimedia Devices (MD) \cite{Dunn2018,Munos2016,Hilty2021,Li2017,lim2018beyond,ballinger2018deepheart}. These sensors and devices are integrated into home objects and infrastructure and connected using network technologies. Each sensor is responsible for one or more task at the same time.

PSNs are used to detect human daily activities and measure conditions in the subject's environment. They can be placed in a living environment or attached to different home objects to detect the subject’s activities such as a sofa, table, bed, chair, or floor containing pressure sensors. BSNs, on the other hand, are used to monitor vital signs and health conditions by measuring physiological parameters and detecting ambulatory activities. Finally, more contextual information related to human activity is collected via MD to monitor movement, environmental changes, and to increase the interaction between the monitored subject and the e-health application. Nowadays, advances in technology have made it possible to use wireless sensors and networked devices, which allow for more accurate and real-time data collection, enabling remote monitoring of patients. This is particularly beneficial for elderly and disabled individuals who may have difficulty traveling to a healthcare facility for regular check-ups. An overview of sensors used in HTMS is depicted in Table \ref{tab:health_monitoring_devices_and_sensors}.

\begin{table}[htbp]
\centering
\begin{tabular}{l|l|l|l}
\hline
Category & Name & Purpose & Data Format \\\hline
PSN & PIR & Motion detection & Categorical \\
& RFID & Persons and objects identification & Categorical \\
& Pressure & Identify location & Numerical \\
& Ultrasonic & Tracking location and posture & Numerical \\
& Contact switches & Open/close door detection & Categorical \\
& Light & Use of light and its intensity & Time series \\
& Temperature & Measure room temperature & Time series \\
& Weight & Elderly weight & Numerical \\
& Humidity & Measure room humidity & Time series \\
& Power & On/off and measure power consumption & Numerical \\
BSN & Accelerometer & Measure acceleration, fall detection, location and posture & Time series \\
& Gyroscopes & Measure orientation, motion detection & Time series \\
& GPS & Motion detection and location tracking & Categorical \\
& ECG & Monitor cardiac activity & Analog signal \\
& EEG & Measure of brain waves & Analog signal \\
& EOG & Monitor eye movement & Analog signal \\
& EMG & Monitor muscle activity & Analog signal \\
& PPG & Heart rate and blood velocity & Analog signal \\
& Pulse oximeter & Measure blood oxygen saturation & Analog signal \\
& Blood pressure & Measure blood pressure & Numerical \\
& SKT & Skin temperature & Numerical \\
MD & Cameras & Monitoring and tracking & Image, video \\
& Microphone & Voice detection & Audio \\
& Speakers & Alerts and instructions & Audio \\
& TV & Visual information & Audio, video \\\hline
        
\end{tabular}
\caption{Health monitoring devices and sensors}
\label{tab:health_monitoring_devices_and_sensors}
\end{table}

PSNs or environmental sensors play a crucial role in capturing and retrieving contextual data about a subject and their surroundings. These sensors can be placed in a living environment or attached to different household objects to detect the subject's activities. For example, a pressure sensor placed on a sofa, table, bed, chair, or floor can detect the subject's movements and daily living activities (ADL). By observing the subject in their environment and their interactions with household objects, a performance level can be measured. For instance, if a motion sensor detects that the user is in the kitchen and a sensor on a cooking object such as a gas stove is turned on, along with water usage or the refrigerator's door being open, it can be inferred that the subject is preparing a meal.

Environmental sensors can provide rich contextual information to detect daily activities and observe human behavior in Health Smart Homes (HSH). Passive infrared (PIR) \cite{franco2010behavioral} sensors are commonly used for detecting a person's presence and ADL \cite{noury2012computer}. Additionally, radio-frequency identification (RFID) technology, which includes both active and passive tags attached to objects and readers worn by the subject, can be used to identify users and objects within the HSH environment \cite{hsu2011novel}. Pressure \cite{jih2006multi} and ultrasonic \cite{hori2005ultrasonic} sensors can also be attached to home objects to track their location and therefore the movements of the user. Contact switches can be used to detect the subject's interactions with other objects in the space, such as a door, window or refrigerator. Environmental sensors can also be used to monitor additional dimensions such as light, temperature, and humidity. They are deployed in different locations to monitor environmental conditions and identify daily activities. Power sensors are used to measure and manage energy consumption and to detect the usage of electrical devices using On/Off events. The usage of electrical appliances such as microwaves, water kettles, toasters, room heaters, and televisions can also be used to detect activity and further refine behavior information. PSNs play a vital role in capturing and retrieving contextual data in health monitoring and smart homes. They can detect daily activities, observe human behavior, and track the movements of individuals. These sensors can also monitor environmental conditions and identify daily activities, measure energy consumption, and detect the usage of electrical devices. Advances in technology have made it possible to use these sensors and networked devices to provide accurate, real-time data collection, enabling remote monitoring of patients.

BSNs utilize wearable sensors that are worn by individuals being monitored, such as the elderly and patients. These sensors are used in Health Monitoring Systems (HMS) to provide continuous information about the individual's real-time health conditions. These sensors are often embedded into everyday accessories such as clothing, belts, watches, or glasses. BSNs often use inertial measurement units, such as accelerometers, to detect ambulatory activities or vital sign devices, such as heart rate sensors, to monitor the individual's health condition. The most commonly used inertial sensors for monitoring movements and body postures are accelerometers and gyroscopes. Accelerometer sensors measure acceleration values and are typically based on three-axis accelerometers positioned on specific locations on the human body. In a study \cite{gao2014evaluation}, the authors used four accelerometers attached to the chest, left under-arm, waist, and thigh to monitor and recognize five activities, such as standing, sitting, lying, walking, and transition.

A study \cite{jiang2011method} proposed a method for recognizing human activities using four wearable accelerometers placed on the forearms and shins of subjects, which monitored ten different daily activities and gym exercises such as standing, sitting, walking, jogging, and weightlifting. Another study \cite{mannini2013activity} used a single accelerometer and gyroscope sensor to monitor 26 events and activities. Gyroscopes are often used in conjunction with accelerometers for movement detection. In \cite{varkey2012human}, a system was proposed that used accelerometer and gyroscope sensors placed on the wrist and foot of the subject to collect linear accelerations and angular rates of motion, which could recognize activities like walking, standing, writing, smoking, jacks, and jogging. Additionally, a fall detection system for the elderly was proposed using accelerometers and gyroscopes \cite{kau2014smart}. Global positioning systems (GPS) can also be used as wearable sensors to monitor location-based activities in open or mobile environments, such as tracking locations or predicting movement across multiple subjects \cite{ashbrook2003using}, or determining mode of transportation based on GPS data logs \cite{liao2007learning}.

In HTMS, various biosensors are used to monitor the vital signs of patients and the elderly, such as heart rate, oxygen saturation, blood pressure, blood glucose, body temperature, and weight. These sensors can be embedded into everyday accessories such as watches \cite{barth2009tempo}, shirts \cite{lee2009wireless}, and belts \cite{kang2007systematic}, and provide real-time physiological information related to the health of the monitored subject. Other biosensors used in HMS include electrocardiography sensors (ECG) for monitoring cardiac activity, electroencephalography sensors (EEG) for monitoring brain activity, electromyography sensors (EMG) for monitoring muscle activity, and electrooculography sensors (EOG) for monitoring eye movements. Pulse oximeters are used to measure the oxygen saturation levels of the blood, while photoplethysmography sensors (EPG) are used to monitor blood flow.

Additional biomedical parameters can also be measured, such as using CO2 gas sensors to evaluate the levels of carbon dioxide in the air to monitor respiration. However, many devices currently available do not have the capability for data relay. Many existing studies and projects have not focused on improving the link between these devices and their infrastructure. The goal is often to transform these sensors into context data sources that can be connected to a network and automatically collect data. While there are various protocols and algorithms that have been proposed for wireless sensor networks, there are currently no features or requirements that are well-suited for body area networks (BANs) \cite{chen2011body}. Factors such as the number of sensors, data rate, mobility, latency, communication, and transmission are determined based on the application and the needs of the subject. Additionally, energy consumption and battery life are still significant challenges for devices in these networks.

The use of home appliances equipped with sensors can help to create an interactive healthcare environment. These appliances can include electrical or electronic devices that serve multiple purposes at home. Examples include cameras, microphones, telephones, speakers, and televisions, which provide a platform for exchanging data between the individual and the system. This increases the individual's interaction with health applications and can serve as new sources for gathering contextual information or as a means for providing guidance and counseling \cite{lemlouma2013smart}.

Multimedia-based approaches involve the use of visual and audio sensing devices such as cameras and microphones. Sound-based detection methods can be used to monitor certain activities of daily living \cite{vacher2011sweet}, while vision-based methods have a broader scope and can be used for posture recognition \cite{brulin2012posture}, detecting human presence \cite{schiele2009visual}, movement and fall detection \cite{taleb2009angelah}, and monitoring complex activities \cite{sacco2012detection}. However, there is debate over the use of multimedia-based applications for healthcare. While these approaches provide rich contextual information, they also have limitations such as computational costs and privacy concerns. The data acquired by these sensors is basic and raw, often imperfect and uncertain in nature, which requires further development to create a high-level approach for healthcare services.

\section{Public Healthcare needs: in the COVID-19 case} 
\label{PublicHealthcareNeeds}
%%Done
 
A fundamental aspect of managing outbreaks is comprehending how the infection spreads in terms of time, place, and individuals, and identifying risk factors for the disease to guide effective interventions. Digital data sources are being used to enhance and interpret key epidemiological data gathered by public health authorities for COVID-19 \cite{Budd2020}.

The COVID-19 pandemic exposed a number of flaws in various systems, such as infrastructure, supply chains, government readiness, human resources, and public health systems. In addition, officials responsible for public health and healthcare system management faced difficulties in conveying a consistent message on public health measures to control the spread of the virus. Healthcare facilities were not properly equipped or prepared to handle the large influx of patients, and medical and epidemiological training was inadequate to provide proper care  \cite{Torre2020}. Overall, public health systems were unprepared to deal with a rapidly spreading viral pathogen, and containment measures were not effectively implemented during the critical period \cite{Torre2020}.

After more than two years since the emergence of the SARS-CoV-2 virus, it is clear that cooperation in sharing information between governments and healthcare institutions, as well as clear and timely communication with the public, is essential to slow the spread of the virus and prevent a resurgence of the pandemic. Nevertheless, it is uncertain whether healthcare measures in any country have adapted sufficiently to cope with future outbreaks \cite{Bhatia2020}.

\subsection{Epidemiological Surveillance}
%done
Traditionally, population-surveillance systems are based on health data from laboratories, notifications of cases diagnosed by doctors, and syndromic surveillance networks. Syndromic surveillance relies on identifying "influenza-like illness" from hospitals and selected primary and secondary healthcare facilities that agree to provide regular surveillance data of all cases. However, these sources often miss cases where individuals do not seek healthcare. For example, in the UK, until recently only hospitalized patients and healthcare workers were routinely tested for COVID-19, meaning that confirmed cases only represent an estimated 4.7\% of symptomatic COVID-19 cases \cite{Unnikrishnan2021}. Identifying undetected cases would help provide a clearer understanding of the outbreak and reduce transmission \cite{Heneghan2020}.

In recent years, data from online news sites, news-aggregation services, social networks, web searches, and participatory longitudinal community cohorts have been used to address this gap. Data-aggregation systems, including ProMED-mail \cite{PROmed}, GPHIN \cite{GPHIN}, HealthMap \cite{HEALTHMAP} and EIOS \cite{EIOS}, which use natural language processing and machine learning to process and filter online data, have been developed to provide epidemiological insight. These data sources are increasingly being incorporated into the formal surveillance landscape \cite{Edelstein2018} and have a role in COVID-19 surveillance. The World Health Organization's (WHO) platform, EPI-BRAIN \cite{epiinfo_2022}, brings together a variety of datasets for infectious-disease emergency preparedness and response, including environmental and meteorological data.

%done 
Some systems have claimed to detect early reports of COVID-19 through the use of crowdsourced data and news reports, before the World Health Organization (WHO) officially announced the outbreak \cite{McCall2020},\cite{PROmed}, \cite{Bogoch2020}. For example, the UK's automatic syndromic surveillance system scans National Health Service digital records \cite{Smith2017} to detect clusters of a respiratory syndrome that could indicate COVID-19. There is also interest in using online data to estimate the true spread of infectious diseases in the community \cite{Gomide2011}, \cite{Lampos2018}.

Research on the epidemiological analysis of COVID-19-related social media content has been conducted for example see \cite{Sun2020}, \cite{Qin2020} and \cite{Lu2020}. Models for COVID-19 \cite{Lampos2021}, which build on previously established internet search algorithms for influenza, are included in Public Health England's weekly reports \cite{Lampos2015}. Crowdsourcing systems used to understand the true burden of the disease are also supporting syndromic surveillance. For instance, InfluenzaNet gathers information about symptoms and compliance with social distancing from volunteers in several European countries through a weekly survey \cite{Koppe2017}. Similar efforts exist in other countries, such as COVID Near You \cite{COVIDNEAR2022} in the USA, Canada, and Mexico. The COVID-19 symptom-tracker app has been downloaded by 3.9 million people in the UK and USA \cite{Menni2020} and is feeding into national surveillance.

\subsubsection{Positive Reporting}
%Done 
The World Health Organization (WHO) has stated that the SARS-CoV-2 virus can be spread through various bodily fluids and respiratory droplets, making personal protective equipment for healthcare workers essential for both treating patients and protecting workers. In the general population, the virus has spread quickly, making testing labs crucial for identifying and tracking the spread of COVID-19. While testing is a reactive measure, it can provide important information about the prevalence of infection and aid in coordinating healthcare for infected individuals \cite{Colson2020}, \cite{Kubina2020}. Standard protocols for dealing with COVID-19 cases often involve transferring patients to designated containment areas, but these protocols vary and there is no cohesive global protocol. This has made it difficult to trace and track patients' contacts and implement population-level containment measures. Even patients who do not require hospitalisation can still pose a risk as they may come into contact with others. Asymptomatic individuals who are not tested can also contribute to the spread of the virus, which is why comprehensive testing strategies are necessary to curb the pandemic.
\begin{figure}[htbp]
    \centering
    \includegraphics[width=0.7\textwidth]{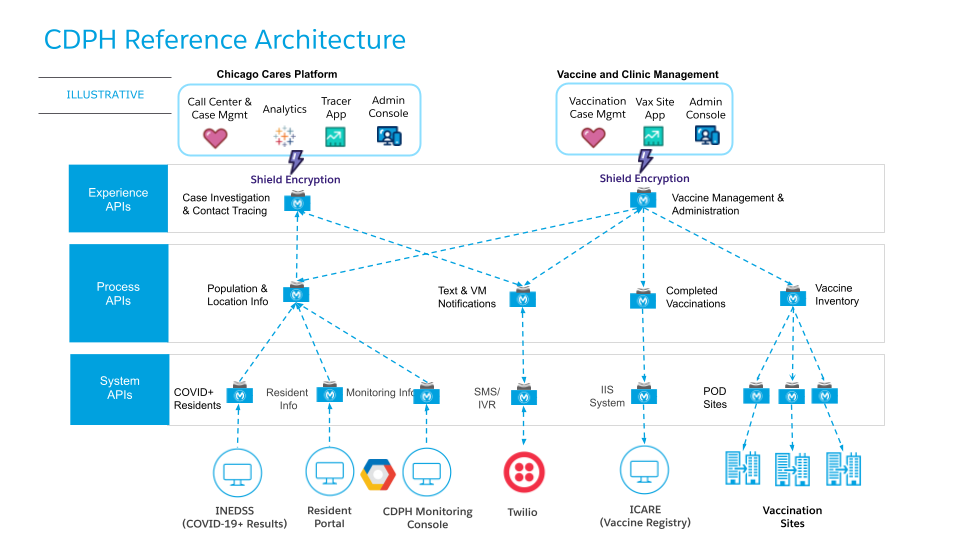}
    \caption{Chicago Department of Public Health COVID-19 Management Architecture \cite{CDPHArchitecture}}
    \label{fig:CDPHReferenceArchitecture}
\end{figure}

%done 
The ability of the virus to spread even when the infected person shows no symptoms is a major obstacle in controlling the COVID-19 pandemic \cite{Christie2021}, \cite{Canadian2020}, and so regular testing of staff working with vulnerable populations is crucial. Rapid antigen tests and at-home testing kits have made it easier to test people, but they do have the potential for false negatives. Nevertheless, these tests and kits are important tools in tracking the spread of the virus.

According to regulations worldwide, laboratories involved in detecting and sequencing the SARS-CoV-2 virus must be managed by trained staff or experts, who must follow strict protocols and use appropriate equipment, including nucleic acid extractors, RT-PCR devices, ultra-low freezers, UV lamps for decontamination, and other disinfection equipment, robots, and contamination-free consumables \cite{Shokoohi2020}, \cite{Gandhi2020}.

\subsubsection{Data analytics} 
%done  
During the pandemic, data dashboards have been widely utilized to collect and display real-time information on public health, including the number of confirmed cases, deaths, and testing results, in order to inform the public and aid decision-makers in adjusting their responses \cite{MHS}, \cite{HongKongHP}, \cite{nextstrain}. These COVID-19 dashboards often present the data in the form of timelines and geographic maps, showing statistics at both the regional and individual level \cite{HongKongHP}, \cite{COVID19SG}. Some dashboards also include information on other aspects of the pandemic, such as clinical trials \cite{Thorlund2020}, government policies, economic measures \cite{WorldBank}, and compliance with social distancing guidelines \cite{DataNetwork}.

%done 
There are few data dashboards that include information on contact tracing or monitoring through apps, and their effectiveness is not well-documented. Problems with the quality and consistency of data collection remain a concern. The absence of official standards and variations in how governments report statistics across different countries make it hard to compare data globally. Additionally, up-to-date and accurate statistics from governments are not always readily available. New ways of visualising data, such as the NextStrain open repository \cite{nextstrain}, which uses viral sequence data to create a worldwide map of the spread of the disease, are emerging. This is made possible by open data sharing and the use of open-source code, which allows for faster sharing of information compared to previous global outbreaks \cite{WHO2018}.

\subsubsection{Vaccination platforms}
%done
A vaccination platform \cite{ward2021digital} is a system that is used to manage the administration, distribution, and tracking of vaccines. This can include features such as:
\begin{itemize}
\item \textbf{\textit{Scheduling}}: Allows individuals to schedule appointments for vaccine doses. This feature allows individuals to select a date, time, and location for their vaccine appointment, and can also include options for rescheduling or canceling appointments.
\item \textbf{\textit{Registration}}: Allows individuals to register and create a profile with their personal information. This can include contact information, medical history, and insurance information. This feature is useful for tracking individuals' vaccination status and ensuring that they receive the correct vaccine.
\item \textbf{\textit{Eligibility}}: Determines whether an individual is eligible to receive a vaccine and provides information on the vaccines that are available. This feature can be based on factors such as age, health condition, and occupation. It can also provide information on the different types of vaccines available and the schedule of doses.
\item \textbf{\textit{Administration}}: Tracks the administration of vaccine doses, including the date and location of the vaccine, and the vaccine type. This feature can also include the option for healthcare providers to enter notes about the vaccine administration, such as any adverse reactions that occurred.
\item \textbf{\textit{Reminders}}: Sends reminders to individuals about upcoming vaccine appointments. This feature can include text messages, emails, or phone calls to remind individuals of their appointment and can also include information on how to reschedule or cancel.
\item \textbf{\textit{Reporting}}: Generates reports on vaccine administration and inventory management. This feature can provide information such as the number of doses administered, the number of doses remaining, and the number of individuals who have received the vaccine.
\item \textbf{\textit{Data Management}}: Stores and manages data related to vaccine administration and inventory. This feature can include options for exporting data to other systems, such as electronic health records (EHRs), and can also include data analytics tools to track the progress of the vaccine distribution.
\item \textbf{\textit{Communication}}: Allows for communication between healthcare providers, government agencies, and individuals. This feature can include options for messaging, email, or phone calls to facilitate communication and ensure that everyone is on the same page.
\item \textbf{\textit{Integration}}: Integrate with other healthcare systems like EHR(Electronic Health Record) to provide a more complete and accurate picture of an individual’s health history.
\item \textbf{\textit{Security}}: Ensures the security and privacy of personal information and vaccine data. This feature can include options for encrypting data and using secure protocols to protect sensitive information.
\item \textbf{\textit{Analytics}}: Provide analytics and insights to monitor the vaccine distribution. This feature can include dashboards and reports to track the progress of the vaccine distribution, such as the number of doses administered, the number of individuals vaccinated, and the number of doses remaining.
\item \textbf{\textit{Multi-lingual support}}: Platform is available in multiple languages to reach more people. This feature can help to ensure that individuals who speak different languages can access the platform and understand the information provided.
\end{itemize}

\begin{figure}[htbp]
    \centering
    \includegraphics[width=0.7\textwidth]{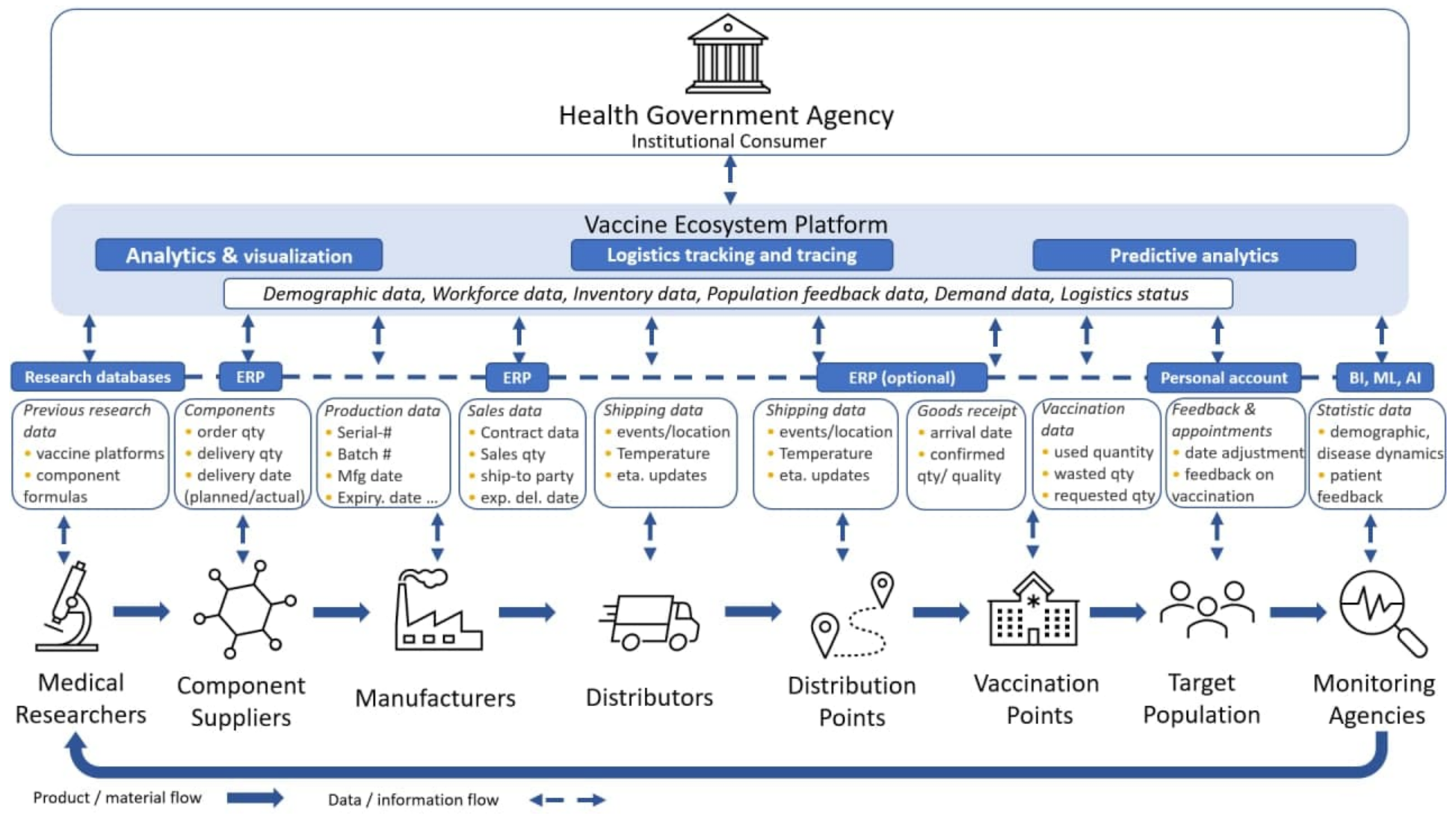}
    \caption{Vaccine ecosystem, Platform Application and Data Architecture \cite{ilin2022innovative}}
    \label{fig:VaccineEcosystemPlatform}
\end{figure}

The platform can be accessed through a website or a mobile application, and can be used by healthcare providers, government agencies, and individuals. The aim of this platform is to simplify the process of managing and tracking vaccines and help to ensure that individuals are able to receive their vaccines as quickly and efficiently as possible.

A decentralized vaccination platform \cite{radonjic2021decentralized} is a system that uses blockchain \cite{NG2021e819} technology to manage the administration, distribution, and tracking of vaccines. In contrast to traditional centralized systems, where a single entity controls and manages the data, a decentralized platform distributes the control and management of the data among multiple participants. It uses a distributed ledger technology (DLT) like blockchain to create a tamper-proof record of vaccine administration and inventory. This allows for secure and transparent tracking of vaccines and ensures that the data is accurate and up-to-date. Some of the features of a decentralized vaccine electronic platform include:
\begin{itemize}
\item \textbf{\textit{Immutable record keeping}}: all the data is stored in a decentralized ledger, making it tamper-proof and resistant to unauthorized modification.
\item \textbf{\textit{Smart contracts}}: the platform can use smart contracts to automate certain actions, such as scheduling appointments or tracking inventory.
\item \textbf{\textit{Anonymized data}}: individuals' personal information can be encrypted and stored on the blockchain, preserving their privacy.
\item \textbf{\textit{Interoperability}}: the platform can be integrated with other healthcare systems like EHR.
\item \textbf{\textit{Access control}}: the platform can be built in a way that allows only authorized parties to access the data.
\item \textbf{\textit{Traceability}}: the platform can provide a complete record of the vaccine administration and inventory, allowing for better traceability and transparency.
\item \textbf{\textit{Public accessibility}}: the platform can be accessible to the public and provide real-time information about the vaccine distribution.
\end{itemize}

The vaccination process spans into three phases \cite{radonjic2021decentralized}:
\begin{itemize}
\item \textbf{\textit{Pre-Vaccination phase}} includes activities in which the knowledge about available vaccines is acquired, the availability of a certain number of vaccines ensured, and patients are selected in terms of scheduling a vaccination appointment.
\item \textbf{\textit{Vaccination phase}} includes activities in which the patient is prepared for vaccination, informed about potential side effects and adverse events, and the vaccination process, including injection, is executed.
\item \textbf{\textit{Post-Vaccination phase}} includes activities in which post-injection monitoring is conducted, and post-procedure administrative tasks are completed. This involves recording vaccination-related data and any side effects that have been occurred during the
vaccination procedure.
\end{itemize}
      
\subsubsection{Clinical Testing platforms}
%done 
Clinical testing platforms existed before the COVID-19 pandemic, but the outbreak put a strain on these platforms, largely due to the high demand and the large number of users accessing them concurrently \cite{testd_2022}. These platforms typically have four main user groups \cite{ixlayer_2022}: individuals who will take the test, lab personnel who process the test and record the results, medical doctors who use the results to monitor patients, and government medical services who use the results to track the epidemiological status of a region. The pandemic led to the creation of several new clinical testing platforms (some of them show in TABLE~\ref{tab:clinical_testing_platforms_features}), or existing platforms being adapted to include clinical testing capabilities. Some platforms were even repurposed as data collection platforms due to the urgent need during the pandemic.

\begin{table}[htbp]
\centering
\begin{tabular}{l|p{13.5cm}}
\hline
Platform & Features \\\hline
DHIS2 \cite{dhis2_2021} & 
    \tabitem Enrolls \& tracks suspected cases, captures symptoms, demographics, risk factors \& exposures, creates lab requests and captures laboratory data about the case, links confirmed cases with contacts and monitors patient outcomes.
    
    \tabitem Strengthens active case detection through contact tracing activities, such as identification and follow-up of contacts of a suspected or confirmed COVID-19 case.
    
    \tabitem Enrolls travelers who have visited high-risk locations at Ports of Entry for 14-day monitoring and follow-up.
    
    \tabitem Simplified line-list that captures a subset of minimum critical data points to facilitate rapid analysis \& response.

    \tabitem Aggregate reporting dataset that captures minimum necessary data points for daily or weekly reporting.
    
\\

SORMAS \cite{sormas_2022}             & 

    \tabitem Enrolls and tracks suspected cases, contact tracing \& follow-up visit, laboratory sample management, event tracking (early warning), patient information system for tracking symptoms.
    
    \tabitem Reporting which includes case count, contact count, lab confirmation and deaths.
    
    \tabitem Statistical analysis based on reports (Maps, charts, graphs, dashboards, etc.). Network transmission chain visualization.
    
    \tabitem Importing files from other entities or datasets and exporting files to transfer data to other entities or datasets.

    \\
Go.Data \cite{GoData_2022}            & 
\tabitem WHO case reporting forms (COVID-MART / X-MART on daily and weekly basis)

\tabitem Generates streamlined Integrated Disease Surveillance and Response (IDSR) reporting 

\tabitem implements COVID-19 First Few Hundred (FFX) Cases Protocol 

\tabitem Case line lists, case investigation protocol, contact line lists, contact follow-up, event \& exposure registration, tracking of relationship \& viewing dynamically chains of transmission
\\

Epi Info \cite{epiinfo_2022}            & 

\tabitem Case Surveillance Forms, custom development to meet local, regional, and country requirements. \\
Open Data Kit (ODK) \cite{odk_2022} & 
    \tabitem Disease surveillance, rapid diagnostics, vaccine trials
    \\
CommCare \cite{CommCare_2022}           & 

\tabitem Fully-functioning application on desktop or mobile browsers and as standalone, offline-capable to carry out disease surveillance and educational activities based on protocol from WHO, CDC, and other leading public health organizations. 

\tabitem Contact tracing using WHO First Few Hundred (FFX) Cases Protocol, Port of entry surveillance, facility readiness and stock tracking, Sample tracking and Lab testing

\tabitem Health Care Provider Training \& Monitoring

\\ \hline
\end{tabular}
\caption{Clinical Testing Platforms Features}
\label{tab:clinical_testing_platforms_features}
\end{table}

      \subsubsection{Decision making}
      %done
     The COVID-19 pandemic has revealed shortcomings in decision-making processes. The rapid spread of the virus within and between countries led to a strain on governments and healthcare organisations. To aid decision-making, various tools, techniques, and strategies have been suggested and implemented. 
      In \cite{Rbbelen2022} they compared static versus interactive decision support tools for COVID-19. They concluded that due to the limited decision time and space during COVID-19, static dicision support tools are more efficient in providing guidance to the interested parties.
      In \cite{McRae2020} they proposed a Clinical Decision Support System (CDSS) and a mobile app to support home care, primary care, urgent care clinics, emergency departments, hospitals and intensive care units. 
      In \cite{Suraj2022} they proposed the SMART COVID Navigator which allows physicians to predict the fatality and severity of COVID-19 progression given a particular patient’s medical conditions. This allows physicians to determine how aggressively to treat patients infected with COVID-19 and to prioritize different patients for treatment considering their prior medical conditions.
      In \cite{Brggemann2021} they proposed an accessible, flexible tool for local health care professionals for a more systematic examination of down-stream effects of decisions under varying and interdependent conditions, to detect a shortage of finite resources such as staff, medication and medical equipment early on.
      In \cite{Saegerman2021} they proposed an interactive and adaptive clinical decision support tool which had a web-based user-interface used to help nurses and clinicians from emergency departments with the triage of patients during the COVID-19 outbrake.
      In \cite{Yang2022} they proposed a toolkit for designers to select design interventions and modelling tools to improve resilience against COVID or future pandemics. To achieve that they had to figure out the pathogen types, transmission modes and phases, and map the design interventions as well as modelling methodologies used. In addition they examined the relationship between different types of infectious diseases.
      In \cite{Silal2022} they describe the adaptation, development and operation of epidemiological surveillance and modelling systems in South Africa in response to the COVID-19 epidemic. They used data from laboratory-confirmed COVID-19 cases, hospitalisations, mortality and recoveries in order to evaluate how these date is used to inform modelling projections and public health decisions. 
      In \cite{Zhigljavsky2020} they modeled and compared different exit scenarios from the lock-down for the COVID-19 UK epidemic. This allowed them to create a modelling basis for laying out the strategy options for the decision-makers.
      In \cite{Chadi2022} they used reinforcement learning to demonstrated that it can be efficient at solving epidemic control problems. They showed that this learning-based approach given the difficulties facing policymakers in defining optimal control policies with different constraints can be useful.
      
\subsubsection{Contact Tracing}
%%done
Contact tracing is crucial in controlling the spread of COVID-19 as it can identify individuals who have been in close contact with confirmed cases, and help enforce quarantine and social distancing measures. By using digital technology for contact tracing, governments can be more efficient and effective in terms of time and resources. However, these digital contact tracing methods raise concerns about data privacy and security, particularly when it comes to location data. While people may be willing to temporarily give up some privacy to fight the pandemic, these concerns may continue in the long term and need to be addressed through technology and policy changes. Failure to address these concerns could lead to a decline in data governance and make it more difficult for governments to gain public cooperation. Additionally, it could cause tension between different groups \cite{Veronica2022}.
\begin{figure}[htbp]
    \centering
    \includegraphics[width=0.7\textwidth]{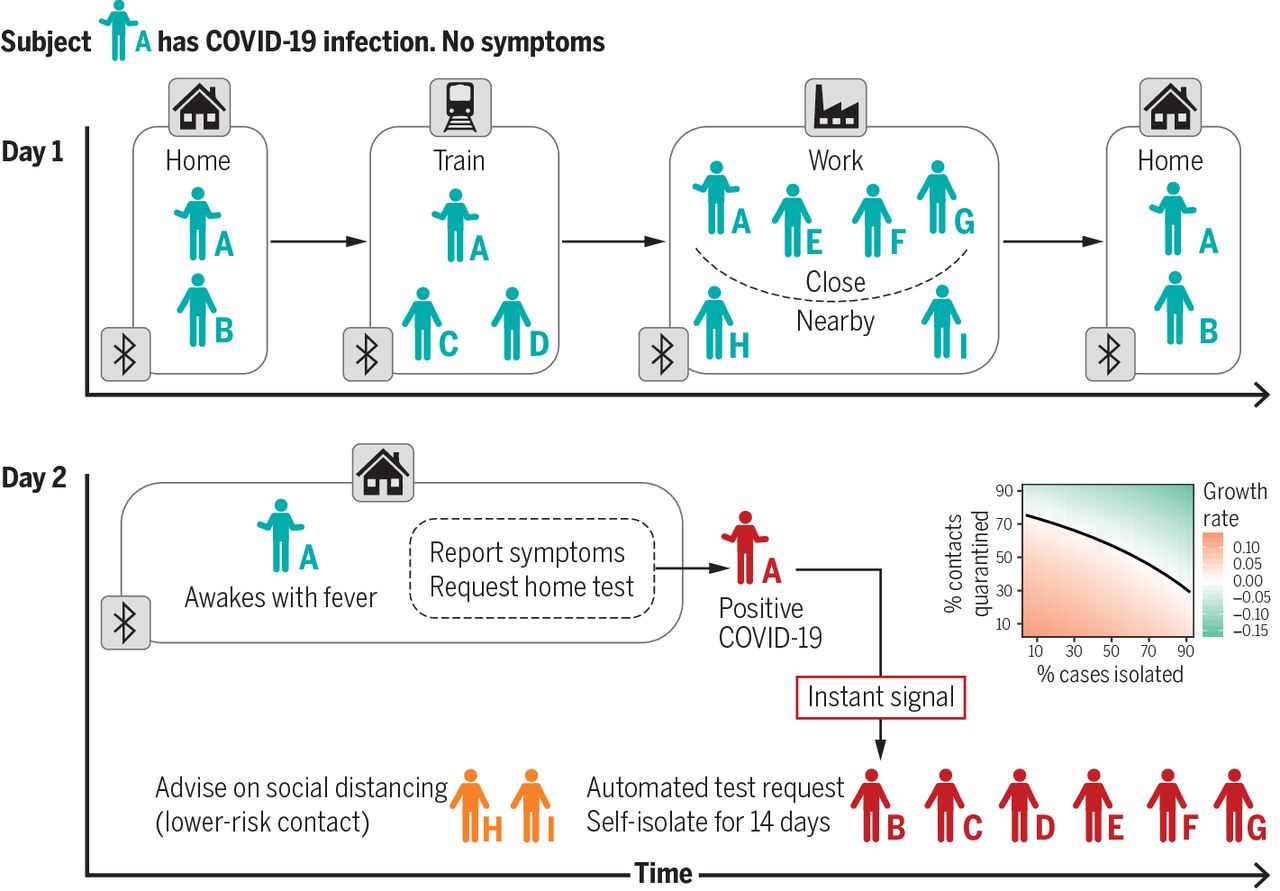}
    \caption{Contact Tracing Impact \cite{ferretti2020quantifying}}
    \label{fig:ContactTracingImpact}
\end{figure}

The incorporation of mobile apps and online platforms in healthcare has greatly expanded patients' access to medical providers. With the widespread availability of healthcare interface applications that connect patients to medical facilities providing real-time results for diagnostic tests and virtual consultations with care providers, the COVID-19 pandemic has highlighted the importance of digital solutions for basic health care issues. The advancement of digital health solutions has significantly transformed the healthcare system by simplifying the process of consulting with medical professionals through telemedicine. Telemedicine offers several benefits such as reducing the risk of infection, eliminating the need for waiting rooms and reducing travel costs, travel time, and lost work time. However, it is not able to replace in-person medical care for certain conditions that require clinical observations, specialized interventions, and exams. \cite{Pabinger2021}.

Digital health applications have revolutionized the healthcare industry by allowing patients to easily access personal health information, diagnostic results, and have virtual consultations with medical professionals. The widespread use of mobile devices has led to an increase in the development and adoption of mobile health care apps, which are becoming more specialized for specific diagnostic tests and can be used with external devices such as inhalers, digital stethoscopes, and blood pressure monitors \cite{Nuttall2021}, \cite{Jamaladin2018}.

The growth of digital health has resulted in over 318,000 medical health apps developed as of 2017 \cite{DigitalHealth2023}, and the mobile health (mHealth) market is expected to reach a value of 111 billion dollars by 2025 \cite{Rowland2020}, \cite{Yan2021}. These technologies are also supported by the U.S. Food and Drug Administration, which has issued the Digital Health Innovation Action Plan in 2017 \cite{DigitalHealth2023} to encourage innovation and regulate digital health.

The Digital Health Innovation Action Plan encourages innovation in digital health and includes three goals: 
\begin{enumerate}
\item providing new guidance for regulating digital health, 
\item developing new methods for oversight, 
\item and increasing expertise within the agency. 
\end{enumerate}

To ensure the safety of medical care, mobile health applications should be regulated and monitored by government health agencies worldwide. In the United States, for instance, mobile health apps are regulated by three federal agencies: 
\begin{itemize}
\item The Office for Civil Rights (OCR) within the Department of Health and Human Services (HHS) enforces the Health Insurance Portability and Accountability Act (HIPAA) rules to protect the privacy and security of health information.
\item The FDA enforces the Federal Food, Drug, and Cosmetic Act (FD\&C Act) to regulate the safety of using medical devices, and eliminate the risk when health app does not work properly. 
\item The Federal Trade Commission (FTC) enforces the Federal Trade Commission Act (FTC Act) to prohibit or create alerts of unfair acts or practices and monitors app's safety or performance.
\end{itemize}

During the COVID-19 pandemic, many governments around the world developed contact tracing mobile alert apps to help control the spread of the virus. The National Health Service in the United Kingdom, for example, developed an app that could warn people when they were in violation of physical distancing guidelines or had been in close contact with a confirmed case. The app was downloaded over 10 million times within ten days of its release. Additionally, the World Health Organization launched a health alert messaging service on WhatsApp that provided the latest information on COVID-19, including symptoms and precautions. This service also provided up-to-date situation reports to aid government officials and health care agencies in taking effective measures to protect public health. As of now, there are about 300 apps that focus on controlling the spread of COVID-19, but their effectiveness varies widely and requires further evaluation \cite{Almaki2021}. As the pandemic continues, more mHealth apps that focus on COVID-19 will be developed and optimized, as they have proven to be useful tools. Examples of mHealth apps used to control the spread of COVID-19 are shown in Table \ref{tab:mhealth_apps}.

\begin{table}[htbp]
\centering
\begin{tabular}{l|l|l|l}
\hline
Country/Region & App & Type of App & Impact \\\hline
Australia \cite{Akinbi2021}, \cite{Singh2020} & COVIDSafe & Tracing & \\ 
&&& Positive -- epidemiological, surveillance,\\
&&& date, time and duration of contact \\
France \cite{Akinbi2021}, \cite{Singh2020} & Stop Covid & &Negative -- user data privacy concerns, lack of trust\\
& TousAntiCovid && ethical issues, security vulnerabilities, technical constraints\\
&&&\\
Germany \cite{Akinbi2021}, \cite{Singh2020} & Corona-Warn-App & & \\
&&&\\
Globally \cite{Indres2021} & TraceTogether & & \\
& Covid Trace &  &\\
& NOVID Share Trace Safe2 & & Positive -- The user receives a unique ID, and at\\
&&& infection, he is redirected to medical applications \\
&&&\\
Italy \cite{Akinbi2021}, \cite{Singh2020} & Immuni & & Positive--Epidemiological surveillance, date, time \\
&&& duration of contact\\
&& &Negative-- user data privacy conscerns, lack of trust, ethical\\
Spain \cite{Akinbi2021}, \cite{Singh2020} & AsistenciaCOVID-19 & & issues security vulnerabilities, technical constraints \\
&&&\\
Switzerland \cite{Akinbi2021}, \cite{Singh2020} & SwissCovid & \\
&&&\\
Cyprus \cite{Isaia2021} & COVTRACER-EN & & \\
&&&\\
\hline
Globally \cite{Indres2021} & HowWeFeel  & Medical & Positive -- for health care professionals and researchers; a useful\\
&&& tool for patients, quarantine advice, and measures to ensure\\
&&& well being. The authorised access was\\
&&& ensured by firewalls, anti-viruses, and cryptographic\\
&&& algorithms\\
&&&\\
Mexico \cite{Morge2022} & Sofia & & Positive--video consultations:internal medicine and \\
&&& pediatric consultations, prescriptions, follow--up indications;\\
&&& high levels of patient satisfaction, versatile and convenient\\
&&& tool to manage the situation\\
&&&\\
USA \cite{Teladoc2022} & Teladoc & & Positive--access to low--cost and high--quality doctors, clinical\\
&&& expertise, virtual care for consumers and clinicians\\
&&&\\\hline
        
\end{tabular}
\caption{Examples of m-Health apps developed by government agencies for contact tracing and information delivery for the spread of COVID--19}
\label{tab:mhealth_apps}
\end{table}

\subsection {Case identification}
  %done
During a pandemic, identifying cases and their contacts quickly is essential for controlling the spread of the virus and understanding how it is transmitted \cite{Budd2020}. The World Health Organisation has stated that while the pandemic has rapidly progressed, the rate of decrease has been much slower \cite{WHO2020}. Therefore, early detection and isolation of those carrying the virus is crucial in preventing further spread of the outbreak \cite{he2020further}, \cite{zhang2020changes}. Early identification of COVID-19 patients can decrease the extent of the virus and reduce the infection and death rate \cite{lai2020severe}, \cite{hellewell2020feasibility}. Studies have demonstrated that NPIs such as monitoring epidemics, investigating epidemiology and tracking close contacts can aid in detecting and reporting potential and confirmed cases promptly \cite{yang2020modified}, \cite{bragazzi2020big}. Big data analysis and artificial intelligence assist in screening for epidemics by collecting personal information, tracking movements and measuring body temperature \cite{xu2020reconstruction}, \cite{bouchnita2020hybrid}, \cite{ting2020digital}, \cite{wang2019effects}. 
\begin{figure}[htbp]
    \centering
    \includegraphics[width=0.7\textwidth]{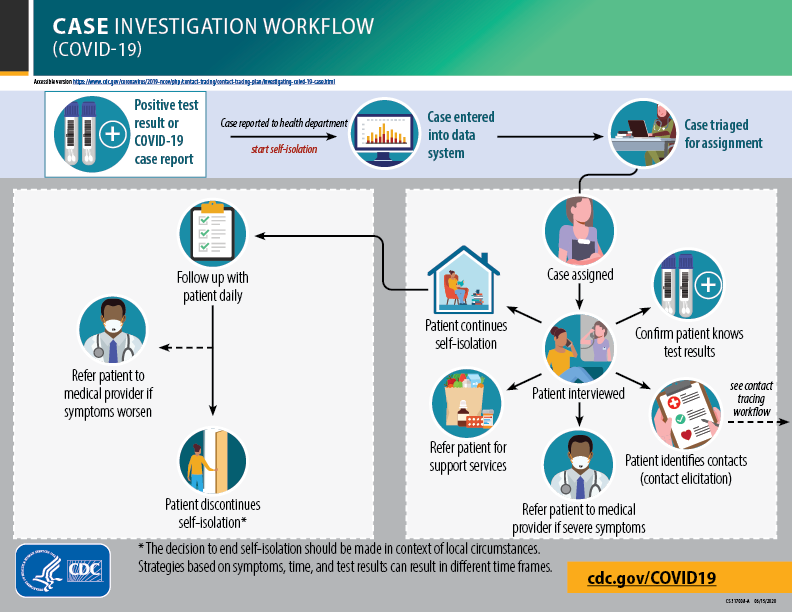}
    \caption{COVID-19 Case Identification \cite{cdcinvestigtingcovidcase}}
    \label{fig:COVID19CaseIdentification}
\end{figure}

In the initial stages of the outbreak, information reporting was the only source of big data, leading to a heavy workload for staff collecting data manually and the possibility of missing important information \cite{chen2020covid}. In China, government departments and internet companies have collaborated to develop various applications to monitor the public's health and individual behavior in real time with the goal of accurately tracking and managing the movement of people \cite{chen2020covid}. Residents report personal information through digital tools, including identification information, daily body temperature, location, and traffic data \cite{lin2020combat}, \cite{he2020accounting}. These tools are based on real data and use advanced algorithms and big data technology to determine if a user is a confirmed case or a potential source of infection \cite{zhu2020analysis}, \cite{wu2020application}.

%Digital technologies can help with this by allowing for symptom-based identification of cases, expanding access to testing, and automating the reporting of test results to public health databases. This can aid in the isolation of cases and their contacts, which is necessary to reduce the spread of the virus.
% \subsubsection{IoT/IoMT}
%\subsubsection{Machine learning}
%\subsubsection{Sensor Wearables}
%done 

Digital technologies can aid in identifying COVID-19 cases by supplementing clinical and laboratory notifications through the use of symptom-based case identification and providing widespread access to community testing and self-testing. This can also automate and speed up reporting to public health databases. Case identification through online symptom reporting, as seen in Singapore \cite{Singapore2020} and the UK \cite{NHS2020}, can provide advice on isolation and referrals to further healthcare services such as video assessments and testing. However, this approach must be linked to ongoing public health surveillance and actions such as isolating cases and quarantining contacts. Widespread testing and contact tracing also play a crucial role in identifying cases, as an estimated 80\% of COVID-19 cases are mild or asymptomatic \cite{Heneghan2020}. Technologies such as sensors, thermal imaging cameras, and wearable devices are also being explored for monitoring COVID-19 in populations \cite{Armitage2020}. 

There has been growing interest in the use of digital rapid diagnostic tests for COVID-19 in order to increase access to testing, boost capacity, and ease the burden on healthcare systems and diagnostic laboratories \cite{Wood2019}, \cite{Land2019}, \cite{Udugama2020}. At the early and mid stage of the pandemic, some point-of-care PCR tests for COVID-19 have been developed \cite{Find2020}, but their use was currently limited to healthcare settings. Drive-through testing facilities and self-swab kits have expanded access to testing, but it was observed delays between sampling, sending samples to centralised labs, waiting for results, and follow-up. On the other hand, rapid diagnostic antibody tests that can be used in home or community settings and provide results in minutes have the potential to improve this process. By linking these tests to smartphones and using image processing and machine-learning methods \cite{Mudanyali2012}, \cite{Med-Tech2020}, mass testing can be linked with geospatial and patient information and quickly reported to both clinical systems and public health systems. However, for this to be successful, standardisation of data and integration of data into electronic patient records are necessary.

%Determining past infections by means of antibody testing is a crucial aspect of monitoring the spread of the disease and assessing the effectiveness of preventive measures such as social distancing. While point-of-care serology tests are available, their accuracy is uncertain, and it is not yet clear how these tests can be used to guide patient care \cite{GOV-UK2020}, \cite{Mallapaty2020}. Additionally, machine learning algorithms are being developed to identify COVID-19 cases by analysing chest scans using computed tomography \cite{Mei2020}, \cite{Wang2020}, \cite{Wynants2020}, but further research is needed to evaluate their effectiveness \cite{Laghi2020}, \cite{Burlacu2020}.

Curbing the spread of COVID-19 requires identifying individuals who have the virus, isolating and treating them, and quarantining their close contacts. To accomplish this, monitoring the health data and activity range of the population in public areas, such as body temperature records, travel trajectories, and close contact records, is crucial \cite{wosik2020telehealth}, \cite{carfi2020persistent}, \cite{menni2020real}. Hence, temperature monitoring in crowded areas such as train stations and subway stations has become an important method for detecting potential cases. However, traditional body temperature detection methods have limitations. In dense scenarios for instance, it requires a lot of resources to use the traditional forehead temperature measurement tools, and large-scale body temperature detection can cause congestion and increase the risk of cross-infection. Infrared thermal imaging temperature measurement systems have therefore been used in public places with high traffic such as airports, stations, and hotels to try and estimate the body temperature at group level \cite{scarano2020facial}, \cite{rane2020design}. However, legacy telecom network capacity is not sufficient for real-time transmission and storage of large amounts of HD image data and dynamic trajectory data. To address this issue, integration of several different tools using 5G technology combined with artificial intelligence, biometrics, and thermal imaging algorithms were employed to quickly measure the temperatures of large populations at multiple railway stations and to identify individuals who were not wearing masks \cite{siriwardhana2020fight}, \cite{pandya2020smart}. This intelligent temperature measurement system not only aimed to prevent cross-infection but also improve detection efficiency.

Artificial intelligence technologies such as face recognition can aid in identifying suspected COVID-19 cases in a timely manner \cite{allam2020artificial}. Traditional access controls such as using turnstiles and punching cards or fingerprints poses a risk of infection through indirect contact. To eliminate this risk, during the pandemic, face recognition access control was widely deployed. Smart cameras with face recognition capabilities were installed at entrances, not only to open and restricted areas but also to track the movement of crowds \cite{wang2021preventing}.

\subsection{Public Communication Platforms}
%done 
The key to successfully carrying out interventions during a pandemic is to educate the public and gain their cooperation through a well-designed communication plan that involves active community involvement to establish trust. With a large number of people having access to the Internet \cite{Telecom2019} and mobile devices \cite{GSMA}, utilising digital platforms for communication can quickly reach a vast audience and mobilise communities. However, there are ongoing challenges such as the spread of potentially false information \cite{Ball2020}, \cite{Depoux2020} and digital disparities \cite{Beauboyer2020}.

The use of online data and social media has been crucial in the way information about the COVID-19 pandemic is being shared with the public \cite{Merchabt2020}. Efforts are being made to counter the spread of false information \cite{WHO2020}, \cite{WHO13} and promote credible sources. For example, Google's SOS alert feature \cite{Google2020} highlights trusted sources such as the World Health Organization in search results. However, there is limited information on the effectiveness of these interventions \cite{Farooq2020}, \cite{Sesagiri2020}, and the definition of misinformation is not always clear \cite{Limaye2020}. Many governments have put COVID-19 information on their websites and are using text messaging to reach people without internet access \cite{united2020}. Chatbots are also being used to provide information and alleviate pressure on non-emergency health advice lines \cite{WhatsApp2020}. Telemedicine, or the use of technology for remote healthcare \cite{Greenhalgh2020}, is becoming more prevalent, particularly in primary care. Digital communication platforms are also helping with social distancing efforts, enabling remote work and classes \cite{anderson2020}, supporting mental health \cite{Liu2020}, and facilitating community mobilization \cite{GoodSAM2020}. However, there are concerns about the security and privacy of these platforms, particularly in regards to the handling of confidential healthcare information.

%  \subsection{Clinical care}
\section{Public Health Emergency response ICT Challenges} 
\label{PublicHealthcareEmergency}

The COVID-19 pandemic has shown that current public health systems need to be revisited and the existing use of technology to fight the pandemic presents numerous challenges. To effectively combat pandemics, it's important to have strong coordination between data, people, and systems at different scales \cite{bardhan2020connecting}. However, public health agencies and healthcare stakeholders often use different systems and data formats, making it difficult to identify trends and create interventions. To improve the response to future pandemics, it's crucial for public health researchers, epidemiologists, and government officials to be connected through integrated systems with shared data. It's also essential to involve people in the fight against COVID-19 by connecting, coordinating, and supporting them through innovative technology.
%done
Digital technologies have been a part of public health strategies for a long time, but have not been adopted as quickly in the public health sector as in other industries. For instance, WHO release guidelines on digital health interventions in 2019 \cite{WHO2018_2}, \cite{who2019}. However, the COVID-19 pandemic has accelerated the development and implementation of new digital technologies. These technologies have the potential to improve epidemic intelligence through online data, detect and track cases and clusters of infections, trace contacts, monitor travel during lockdowns, and disseminate public health messages on a large scale. Despite these potential benefits, there are still barriers to the widespread use of digital solutions in public health.

%done 
The effectiveness of big data and AI methods depend on the quality of the data that is inputted into them. However, public health and private data sets that are necessary for research are often not easily accessible due to privacy and security concerns \cite{Oliver2020}, \cite{Buckee2020}, \cite{McKendry2020}. These datasets also lack standardization and completeness. Researchers are urging technology and telecommunications companies to share their data in an ethical and privacy-respectful way, citing a moral responsibility for these companies to contribute when there is a valid cause for using the data \cite{ChinaDataLab}, \cite{Google2020}, \cite{Pham2020}, \cite{Microsoft2020}. Some companies are already making aggregated data available, but further steps need to be taken in order to make this data consistent and timely, and there is no standard format or long-term commitment.

%done
Researchers have been working together globally to gather data from various sources that have been voluntarily reported \cite{Mobility}. Additionally, governments should be more transparent and make their data, including information on epidemiology and how the disease spreads, easily accessible to researchers in downloadable formats. Some governments have made individual-level data available for research \cite{Korea2020}, \cite{Government2020}, but this raises concerns about privacy. The use of open-source data, code, and scientific methods is becoming more common, with many being shared online, including the use of pre-prints which makes data available faster but hasn't undergone peer review  \cite{Kupf2020}. Nevertheless solid foundations have been set in the EU\footnote{https://covid19scenariohub.eu/} and in the USA\footnote{https://covid19scenariomodelinghub.org/}, respectively, to facilitate the collective gathering of relevant epidimiological data and collaborating in developing models for analysis and prediction of future patterns.

The impact of COVID-19 has been widely discussed in various papers. Chakraborty and Maity \cite{chakraborty2020covid} looked at the impact of COVID-19 on society and the environment and suggested preventative measures such as controlling population growth, avoiding large gatherings, and developing a safe vaccine. They did not mention the use of IoT as a preventative strategy. Oyeniyi et al. \cite{oyeniyi2020application} focused specifically on the use of IoT in the medical field to treat COVID-19 patients remotely by collecting data through smartphone apps. Nasajpour et al. \cite{nasajpour2020internet} discussed the various applications of IoT in different phases of COVID-19, such as early diagnosis, quarantine, and recovery. Mondal et al. \cite{mondal2021role} presented a survey on the role of IoT in fighting COVID-19 and proposed an IoT-based framework for early detection. Kumar et al. \cite{kumar2021application} proposed a smart healthcare system using IoT to track quarantine patients and provide remote health monitoring. Malliga et al. \cite{malliga2021comprehensive} explored the use of technology to maintain social distancing. Bassam et al. \cite{al2021iot} proposed a three-layer IoT infrastructure for remote health monitoring. Singh et al. \cite{singh2020internet} explored the use of IoT in orthopedic treatment for patients in remote locations. The pandemic has also impacted business and supply chains, leading researchers to focus on supply chain optimization. Chowdhury et al. \cite{chowdhury2021covid} reviewed the effectiveness of supply chain measures during the pandemic. Shahed et al.\cite{shahed2021supply} developed a three-layer optimization model to improve inventory and reduce supply chain disruptions. De Sousa Jabbour et al. \cite{de2020sustainability} discussed the role of supply chain managers in building a smarter and more resilient supply chain management system. Karmaker et al. \cite{karmaker2021improving} proposed a methodology to improve supply chain sustainability in the context of COVID-19. The impact of COVID-19 on the food and beverage industry was discussed by Chowdhury et al. \cite{chowdhury2020case}, who analyzed the short and long-term effects and suggested strategies for recovery such as reducing operational expenses and restructuring the supply chain.

Overall, during and after the pandemic, public health emergency response teams face a number of challenges when it comes to using information and communication technologies (ICTs) to respond to the aftermath. Some of the key challenges posed by the pandemic COVID-19 \cite{he2021information}:
\begin{enumerate}
    \item Overwhelming demand for healthcare services: The pandemic has resulted in a sudden surge in demand for healthcare services, with hospitals and clinics facing an enormous burden to provide care to COVID-19 patients while also attending to other patients with non-COVID related conditions.
    \item Limited availability of medical resources: The pandemic has exposed existing shortages in the medical resource sector, including a lack of medical supplies, personal protective equipment (PPE), and healthcare workers.
    \item Telehealth and remote monitoring: The need for social distancing measures has led to a rapid increase in tele-health and remote monitoring solutions, but these solutions require access to reliable internet and appropriate equipment, which may not be available for all patients.
    \item Integration and interoperability of health data systems: The need for real-time and accurate health data sharing has exposed the limitations of existing health data systems, particularly in terms of integrating and exchanging data between different systems.
    \item Cybersecurity and data privacy: The increased use of digital health technologies has heightened concerns over cybersecurity and data privacy, particularly in the context of sensitive personal and medical information.
    \item Digital divide and access to technology: The pandemic has amplified existing disparities in access to technology and digital health services, with marginalized and low-income communities at higher risk of being left behind.
\end{enumerate}

Those challenges were mostly addressed by the integration of new technologies like the IoT, big data analytics, AI, and blockchain. Moreover, this kind of technologies is being used to find solutions to problems caused by the coronavirus. For instance, Facebook has used AI and big data to map population density, demographics, and travel patterns to determine where to send supplies and minimize the spread of the virus \cite{holt2020facebook}. Big data analysis of GIS and IoT data from infected patients can assist in tracing the source of the virus and identifying close contacts \cite{he2020using}. The U.S. National Science Foundation has funded a project that explores the use of social media big data, geospatial data, and AI for spatial epidemiology research and risk communication. The combination of blockchain, IoT, and AI has potential in addressing trust and security in public health, by storing medical device data and non-personal sensor data on the blockchain while keeping patients' personal data private \cite{gurgu2019does}, \cite{singh2020blockiotintelligence}. AI and big data technologies can be utilized to analyze both on-chain and off-chain data and provide real-time analytics and recommendations through customized dashboards.

The COVID-19 pandemic has shown the necessity of transforming the public health system from being reactive to proactive and creating innovations that can give timely information for proactive decision-making in local, state, and national public health systems \cite{rai2020editor}. Therefore,  pandemic has brought about significant changes to the way we live, work, and interact with one another. In the post-pandemic world, many sectors are likely to see a greater reliance on technology to continue with business as usual. IoT (Internet of Things) technologies can play a crucial role in facilitating the transition to the new normal by enabling remote work, telehealth services, and online learning. By connecting devices and collecting data, IoT can help monitor and manage the spread of the virus, improve access to healthcare services, and provide efficient and effective remote work solutions. Additionally, IoT technologies can drive efficiency, productivity and innovation across various industries, thus boosting the economy. The post-COVID world presents an opportunity to embrace and harness the power of IoT to create a more connected, resilient and sustainable society.Therefore, some of the key challenges that we are facing after the pandemic and need to be addressed includes \cite{mondal2022role}:
\begin{enumerate}
    \item Sustaining telehealth services: The pandemic has led to a significant increase in the use of telehealth services, but maintaining these services in the long term will require addressing issues of reimbursement, interoperability, and data privacy.
    \item Addressing the digital divide: Ensuring equitable access to technology and digital health services for all communities will be a critical challenge in the post-pandemic period, particularly in light of the existing disparities that have been exacerbated by the pandemic.
    \item Integration of digital health solutions into existing healthcare systems: The rapid adoption of digital health solutions during the pandemic has highlighted the need for these solutions to be integrated into existing healthcare systems in a seamless and efficient manner.
    \item Data privacy and security: Ensuring the security and privacy of patient health data will remain a critical challenge in the post-pandemic period, particularly given the increasing amount of sensitive medical information being generated and stored digitally.
    \item Personalized and predictive healthcare: The increased use of digital health technologies has the potential to drive personalized and predictive healthcare solutions, but this will require overcoming challenges related to data sharing and integration, privacy, and cybersecurity.
    \item Addressing workforce needs: The post-pandemic period will require addressing the needs of the healthcare workforce, including ensuring the appropriate training and reskilling to make effective use of new digital health technologies.
    \item Return to normal: As society returns to normal, ICTs will be important in tracking and responding to any resurgence of the disease, as well as in providing information and support to individuals and communities as they adjust to the new normal.
\end{enumerate}

Overall, effective public health emergency response during a pandemic requires a coordinated and comprehensive approach to addressing these ICT challenges, including improving access to technology, providing clear and accurate information, protecting personal privacy, promoting interoperability, and strengthening cybersecurity. Moreover, addressing these ICT challenges after a pandemic requires a continued coordinated and comprehensive approach, including effective data management, telemedicine, vaccination programs, mental health support and contingency plans for future outbreaks.

\section{Discussion}
\label{Discussion}

\subsection{Electronic Health Records (EHR)}

EHRs are digital versions of the traditional paper-based medical records that healthcare providers use to store patient information. EHRs have several advantages over traditional records, including improved data security, increased efficiency, and better patient care. One of the main benefits of EHRs is improved data security. Electronic records are stored on secure servers, which can be accessed only by authorized personnel. This eliminates the risk of lost or stolen paper records, and ensures that patient information is kept confidential.

Another advantage of EHRs is increased efficiency. With electronic records, healthcare providers can access patient information quickly and easily, which can lead to faster diagnoses and treatment. Electronic records also make it easier for healthcare providers to share patient information with other providers, which can improve communication and coordination of care. EHRs can also improve patient care by providing healthcare providers with more accurate and complete patient information. Electronic records can include a wide range of data, such as lab results, X-ray images, and medication lists. This can help providers make more informed decisions about patient care and treatment. In conclusion, Electronic Health Records (EHRs) have many benefits over traditional paper records. They improve data security, increase efficiency, and better patient care. EHRs are becoming more prevalent in the healthcare industry, as more and more healthcare providers are adopting them to improve the quality of care they provide to patients.

\subsection{Hospital Management Systems (HMS)}

HMS are a complex and vital aspect of healthcare, but they come with a number of challenges \cite{kim2012human,mendez2010hospital,cazzaniga2015uses,clark2004challenges}. High cost is one of the main challenges, as both the development and deployment of the system can be expensive. Additionally, the vast number of features in HMS can make the design of the system complex, which can lead to difficulties with the user interface and user experience. Data security is another major challenge, as healthcare organizations must ensure that patient information is kept confidential and in compliance with changing regulations.

Another challenge is migrating from existing manual processes to an HMS. This can be difficult for both staff and patients, especially individuals who may not be familiar with computer technology. Additionally, patients can become impatient during the transition period as they may not be used to the new system.

Despite these challenges, there are several advantages to using an HMS. One of the main advantages is time-saving, as technology can automate many manual tasks. Additionally, HMS can improve efficiency. Furthermore, HMS can be cost-effective and easily manageable, providing easy access to patient data and history. It can also improve patient care, monitor inventory and reduce the workload of documentation. Lastly, it can provide better audit controls and policy compliance.

\subsection{Health Tracking \& Monitoring Systems}

There are several limitations to the use of wearable technology in healthcare. High device cost is a concern that may hinder the accessibility of this consumer healthcare technology to benefit only those who can afford the devices. Furthermore, many devices are not compatible across platforms (e.g., iOS and Android smart phone operating systems) and there is a lack of data standards which limits their broad use and ubiquity. The time needed to choose and properly set up a device, and remembering to charge the device are also potential barriers to proper use. The acceptability of these devices and potential stigma associated with them varies by community and may also be a factor in dividing who benefits from these technologies \cite{farrington2016wearable}.

Most importantly, the accuracy of the devices and evidence that they do in fact improve health outcomes must be established with large clinical and field trials. It is common to find unsubstantiated scientific claims with no reference to peer-reviewed studies on commercial wearable device company websites and advertising materials that can set a dangerous precedent if inaccurate or misinterpreted data is factored into medical decision making. False negatives can cause a potentially fatal condition to be missed while false positives can lead to over-treatment and/or anxiety. Inaccuracy of activity trackers may also lead individuals to overestimate their level of physical activity, limiting their effectiveness for lifestyle interventions \cite{wallen2016accuracy,dooley2017estimating}. \cite{shcherbina2017accuracy} demonstrated that most wrist-worn devices adequately measure HR in laboratory-based activities, but poorly estimate energy expenditure and caution the use of those measurements as part of health improvement programs. \cite{leth2017evaluation} demonstrated that step counts are most accurate at slow walking speeds. While useful for just-in-time interventions, real-time delivery of results should be carefully considered to ensure that accurate data is presented at the right way and at the appropriate time to avoid potentially harmful interventions and patient confusion and anxiety. Training should be provided to health practitioners to increase awareness of the utilities and limitations of wearable technologies \cite{piwek2016rise}. New tools should be developed to assist with the interpretation of wearable device data in a clinical setting. Finally, it is important to further evaluate the emotional burden and increased distraction and technology addiction that wearables can inflict, potentially causing more harm than good.

New and extremely novel wearable sensors are being developed that build upon movement, physiological and biochemical sensors in existence to improve disease detection and prognostication through changes in physiology over time or as a result of a specific treatment or intervention. Mechanical properties of soft tissues were recently monitored via wearables to determine spatiotemporal changes in viscoelasticity of basal cell carcinoma lesions \cite{dagdeviren2015conformal}. Wearable oxygen diffusion sensors have been used to track wound healing, and moisture sensors can inform better timing of wound dressing changes \cite{niederauer2017prospective, milne2016wearable}.

\subsection{Epidemiological Surveillance}

Epidemiological surveillance of infectious viruses including COVID-19 is the process of identifying, tracking, and monitoring cases of the disease in order to understand its spread and inform public health response efforts. There are several key challenges to effective epidemiological surveillance \cite{ibrahim2020epidemiologic}, including:
\begin{enumerate}
    \item Identification and tracking of cases: Identifying and tracking all cases can be difficult, especially in areas with limited testing capacity or where asymptomatic cases are prevalent.
    \item Data quality and completeness: Ensuring that data collected on cases is accurate and complete is important for effective surveillance, but this can be difficult in areas with limited resources or where data systems are not fully developed.
    \item Data sharing and dissemination: Sharing data on positive cases in a timely and transparent manner is important for effective surveillance and response, but this can be challenging in situations where data sharing is hindered by legal or political barriers.
    \item Privacy and confidentiality: Protecting the privacy and confidentiality of individuals is an important consideration in surveillance efforts, but this can be difficult to achieve in practice.
    \item Limited resources: Limited resources, including funding, staff, and equipment, can make it difficult to conduct effective surveillance.
    \item Public trust: Building public trust in surveillance systems and ensuring that surveillance data is perceived as reliable and trustworthy is crucial to effectively track and respond to the pandemic.
\end{enumerate}

Overall, epidemiological surveillance of infectious viruses is a critical tool for understanding the spread of the disease and guiding public health response efforts. However, effectively addressing the challenges to surveillance is crucial for ensuring that surveillance data is accurate, complete, and actionable.

\subsection{Case identification}

Case identification is a critical component of the response to epidemics/ pandemics. It involves identifying individuals who test positive  and isolating and treating individuals, as well as quarantining their close contacts. This is crucial in controlling the spread of the virus. There are several methods used for case identification, including:
\begin{enumerate}
    \item Testing: Testing is the most widely used method for identifying cases. It involves taking samples from individuals and analyzing them for the presence of the virus.
    \item Symptom-based identification: Individuals who have symptoms consistent with the virus signatures, such as fever, cough, and difficulty breathing, may be considered as cases even if they have not been tested.
    \item Contact tracing: Close contacts of known positive cases may be considered as cases even if they have not been tested or have no symptoms.
    \item Digital tools: Government departments and internet companies have jointly developed digital tools in some countries that monitor the health of the public, personal information and behavior in real-time, with the goal of accurately tracking and managing the flow of people to help identify potential cases.
    \item Artificial intelligence and biometrics: AI and biometrics such as face recognition and thermal imaging technology can be used to identify potential cases in crowded places such as airports and train stations.
\end{enumerate}

Case identification is an essential part of controlling the spread of infectious viruses, it can be done through a combination of different methods, and different methods may be more effective in different situations and settings.

\subsection{Public Communication Platforms}
Public communication platforms play a critical role in providing accurate and timely information to the public. They can be used to disseminate information, the viral transmission pathways, and preventative measures, as well as to provide updates on the number of cases and deaths, and the status of the response efforts.
\begin{enumerate}
    \item Official websites: Governments and health organizations often have official websites dedicated to providing information on the pandemic. These websites can be a reliable source of information on the virus, and can be updated in real-time.
    \item Social media: Social media platforms like Twitter, Facebook, and Instagram are widely used to disseminate information on the pandemic. These platforms can be used to share official information, as well as to share personal experiences and perspectives on the virus.
    \item News outlets: News outlets such as television, radio and newspapers can provide information on the pandemic, as well as updates on the response efforts.
    \item Public health hotlines: Governments and health organizations may establish public health hotlines that the public can call for information on the pandemic.
    \item Mobile apps: Some governments have developed mobile apps that provide information on the pandemic, including the number of cases, locations of testing centers, and other relevant information.
    \item Public awareness campaigns: Public awareness campaigns can be used to educate the public on how to prevent the spread of the virus and what to do if they suspect they have been exposed.
\end{enumerate}

Effective public communication during a pandemic requires a multi-channel approach, utilizing different platforms to reach a diverse audience with accurate and timely information. It is important that the information provided is clear, consistent, and easy to understand in order to avoid confusion, fear, and mistrust. Additionally, it is important to have a consistent message across all communication channels in order to avoid conflicting or confusing information being disseminated to the public.

\subsection{Public Health Emergency Response (ICT)}

The COVID-19 pandemic has highlighted the importance of information and communication technology (ICT) in public health emergency response. ICT has been crucial in managing the outbreak, including in areas such as surveillance, contact tracing, and communication with the public. However, the pandemic has also exposed a number of challenges related to the use of ICT in public health emergencies. One of the key challenges has been data management and sharing. With a large volume of data being generated by testing and surveillance, it has been difficult to effectively manage, process, and analyze this information. This has led to delays in identifying and responding to outbreaks, as well as difficulties in tracking the spread of the virus. Additionally, there have been concerns about data privacy and security, particularly when it comes to the sharing of personal information. Another challenge has been the use of digital tools for contact tracing. While these tools can be effective in identifying and isolating individuals who have been in contact with infected individuals, there have been concerns about the accuracy of the data and the potential for privacy violations. Furthermore, there has been a lack of standardization in the use of these tools, which has led to confusion and difficulties in integrating data across different systems. Communication with the public has also been a challenge during the pandemic. There has been a significant amount of misinformation circulating online, which has made it difficult for the public to distinguish between credible and unreliable sources of information. Furthermore, there has been a lack of coordination between different levels of government and between different countries, which has led to confusion and conflicting messages. Another challenge has been the digital divide. While ICT has been crucial in managing the pandemic, not everyone has equal access to these tools. This has led to disparities in the ability to access information and participate in contact tracing and other digital health initiatives. For example, older adults, low-income individuals, and people in rural areas may have limited access to digital devices or internet connectivity. Finally, there have been issues related to cybersecurity during the pandemic. The increased use of ICT has led to an increase in cyber attacks, particularly targeting healthcare organizations. This has led to concerns about the security of personal and sensitive information, as well as the potential disruption of critical services. In conclusion, the COVID-19 pandemic has highlighted the importance of ICT in public health emergency response. However, it has also exposed a number of challenges related to data management and sharing, contact tracing, communication with the public, the digital divide, and cybersecurity. Addressing these challenges will be crucial in effectively managing future public health emergencies.

\section{Conclusions}
\label{Conclusions}

The COVID-19 pandemic has been a significant global event that has affected nearly every aspect of society. It has highlighted a number of important lessons that can be learned in order to better prepare for and respond to future pandemics. The most important lessons are: 
\begin{enumerate}
    \item Early detection and rapid response are crucial in containing outbreaks. Identifying and containing outbreaks early on through better surveillance systems and rapid testing capabilities is imperative in preventing the rapid spread of diseases.
    \item International cooperation and coordination are essential in combating pandemics. As the pandemic has shown, a coordinated global response is necessary to effectively contain and control the spread of diseases that do not respect borders.
    \item Clear and consistent communication is vital during pandemics. The timely and accurate dissemination of information is critical in helping the public understand the risks and take necessary precautions to protect themselves and their communities.
    \item The pandemic has shown the importance of investing in public health systems and emergency preparedness. The pandemic has exposed weaknesses in many healthcare systems and has highlighted the need for better preparedness and planning for future crises.
    \item Digital technology and telemedicine are valuable tools in the fight against pandemics. The pandemic has accelerated the adoption of digital solutions, such as remote consultations and treatment, which have proven to be valuable in the fight against the spread of diseases.
    \item The pandemic has emphasized the importance of equity and social determinants of health. It has disproportionately affected marginalized communities and has highlighted the need to address underlying issues of inequality and create a more equitable society.
\end{enumerate}

\section*{Acknowledgments}
This work was supported by the European Union's Horizon 2020 research and innovation programme under grant agreement No 739551 (KIOS CoE - TEAMING) and from the Republic of Cyprus through the Deputy Ministry of Research, Innovation and Digital Policy. It was also supported by the CIPHIS (Cyprus Innovative Public Health ICT System) project of the NextGenerationEU programme under the Republic of Cyprus Recover and Resilience Plan grant agreement C1.1l2.

\bibliographystyle{IEEEtran}
\bibliography{paper}
\end{document}